%% file: main.tex
\documentclass[a4paper,11pt]{article}
\usepackage{jheppub}
\allowdisplaybreaks[1]

\def\beq{\begin{equation}}
\def\eeq{\end{equation}}
\def\beqa{\begin{eqnarray}}
\def\eeqa{\end{eqnarray}}
\newcommand{\nn}{\nonumber}

\def\eqn#1{Eq.~(\ref{#1})}
\newcommand\Eqns[2]    {Eqs.\,(\ref{#1}) and~(\ref{#2})}

\newcommand\sss{\scriptscriptstyle}
\newcommand\as{\alpha_{\sss S}} 
\newcommand\gs{g_{\sss S}}

\newcommand{\la}{\langle}
\newcommand{\ra}{\rangle}

\newcommand{\ord}{{\cal O}}

\newcommand{\yh}{y_{\rm H}}
\newcommand{\ph}{p_{\rm H}}
\newcommand{\mh}{m_{\rm H}}

\newcommand{\qt}{q_{\sss\perp}}
\newcommand{\pat}{p_{3\sss\perp}}
\newcommand{\pht}{p_{{\rm H} \sss \perp}}
\newcommand{\phih}{\phi_{\rm H}}
\newcommand{\php}{p_{\rm H}^+}
\newcommand{\phm}{p_{\rm H}^-}
\newcommand{\mht}{m_{{\rm H} \sss \perp}}
\newcommand{\sah}{s_{3{\rm H}}}

\def \Tr {\mbox{Tr\,}}

\def\eps{\varepsilon}

\title{The two-loop Higgs impact factor}

\author[a]{Vittorio Del Duca,}
\author[b,c]{Giulio Falcioni}

\affiliation[a]{INFN Laboratori Nazionali di Frascati, 00044 Frascati (Roma), Italy}
\affiliation[b]{Physik-Institut, Universit\"{a}t Z\"{u}rich, Winterthurerstrasse 190, 8057 Z\"{u}rich, Switzerland}
\affiliation[c]{Dipartimento di Fisica, Universit\`{a} di Torino, and INFN Sezione di Torino, Via Pietro Giuria 1, I-10125 Torino, Italy}

\emailAdd{delduca@lnf.infn.it}
\emailAdd{giulio.falcioni@unito.it}

\abstract{In the HEFT, we consider the Regge limit of the two-loop amplitudes for Higgs boson production in association with a jet, expanded to NNLL accuracy.
We discuss the issue of the Regge cuts versus poles in this context, showing that the former cannot contribute through three loops, due to the simplicity of the colour structure of the amplitudes. Finally, we determine for the first time the Higgs impact factor at two-loop accuracy.}

\preprint{~}

\begin{document}
\noindent

\preprint{\begin{flushright}
    ZU-TH 26/25
\end{flushright}}

\maketitle
\flushbottom

\section{Introduction}
\label{sec:intro}

The Regge limit~\cite{Regge:1959mz} of $2 \to 2$ scattering amplitudes is defined as the limit in which the squared center-of-mass energy $s$ is much larger than the momentum transfer $t$.
In the Regge limit, any $2 \to 2$ scattering process is dominated by the exchange in the $t$ channel of the highest-spin particle. In the case of QCD or ${\cal N} = 4$ Super Yang-Mills (SYM) theory, that entails the exchange of a gluon in the $t$ channel. Contributions that do not feature gluon exchange in the $t$ channel are power suppressed in $t/s$. 

In the Regge limit, virtual radiative corrections to the $2 \to 2$ scattering amplitudes and radiative emissions, i.e. $2 \to n$ amplitudes with $n\ge 3$, display universal, i.e. process-independent, features, related to the ordering in rapidity of the outgoing particles.
Building upon that, the Balitsky-Fadin-Kuraev-Lipatov (BFKL) equation resums the radiative corrections to parton-parton scattering to all orders at leading logarithmic (LL)~\cite{Lipatov:1976zz,Kuraev:1976ge,Kuraev:1977fs,Balitsky:1978ic} and next-to-leading logarithmic (NLL) accuracy~\cite{Fadin:1998py,Ciafaloni:1998gs,Kotikov:2000pm,Kotikov:2002ab} in $\log(s/|t|)$. The resummation of large energy logarithms allows for the description of scattering events where jets with large rapidity intervals are detected~\cite{Mueller:1986ey,DelDuca:1993mn,Stirling:1994he,Bartels:1996gr,Andersen:2001kta,Colferai:2010wu,Andersen:2011hs,Ducloue:2013hia,Celiberto:2015yba,Celiberto:2020wpk}. Further, the Regge limit has been explored extensively in both ${\cal N} = 4$ SYM~\cite{Bartels:2008ce,Dixon:2012yy,Basso:2014pla,DelDuca:2016lad,DelDuca:2018hrv,Marzucca:2018ydt,DelDuca:2019tur,Caron-Huot:2020vlo} and QCD~\cite{Fadin:1989kf,Fadin:1993wh,Fadin:1994fj,DelDuca:1995zy,DelDuca:1995ki,Fadin:1995xg,Fadin:1995km,Fadin:1996nw,DelDuca:1996nom,Fadin:1996tb,Fadin:1996yv,DelDuca:1998kx,DelDuca:1998cx,Bern:1998sc,DelDuca:2001gu,Caron-Huot:2013fea,Caron-Huot:2017fxr,Caron-Huot:2017zfo,Fadin:2017nka,Caron-Huot:2020grv,Caola:2021izf,Falcioni:2021buo,Abreu:2024mpk,Buccioni:2024gzo,Abreu:2024xoh} amplitudes and cross sections~\cite{DelDuca:2013lma,DelDuca:2017peo}, and it has been used to constrain, compute or validate amplitudes in general kinematics~\cite{DelDuca:2009au,DelDuca:2010zg,Dixon:2014iba,Henn:2016jdu,Caron-Huot:2016owq,Almelid:2017qju,Caron-Huot:2019vjl,Falcioni:2021buo,Caola:2021izf}.
Finally, the Regge limit allows also for an effective field theory description~\cite{Lipatov:1995pn,Balitsky:1995ub,Balitsky:1998ya,Kovchegov:1999yj,Kovchegov:1999ua,Jalilian-Marian:1997qno,Jalilian-Marian:1998tzv,Iancu:2000hn,Ferreiro:2001qy,Chiu:2011qc,Chiu:2012ir,Caron-Huot:2013fea,Rothstein:2016bsq,Moult:2022lfy,Gao:2024qsg,Rothstein:2024fpx}.

The BFKL equation describes the rapidity evolution of the gluon ladder exchanged in the $t$ channel in terms of an integral over transverse momentum. The logarithmic accuracy of the equation is driven by the accuracy of its kernel. At LL accuracy~\cite{Lipatov:1976zz,Kuraev:1976ge,Kuraev:1977fs,Balitsky:1978ic}, the leading-order kernel is composed by the central-emission vertex (CEV) of the gluon~\cite{Lipatov:1976zz}, which first occurs in the tree-level $2\to 3$ amplitude in multi-Regge kinematics (MRK), where the outgoing partons are strongly ordered in rapidity.
The soft divergences of the kernel are regulated by the soft structure of the one-loop gluon Regge trajectory, which arises from the virtual radiative corrections to the $2 \to 2$ scattering. 
At LL accuracy, each gluon emission along the ladder introduces a factor of ${\cal O}(\as \log(s/|t|))$ after rapidity integration, thus the BFKL equation resums the corrections of ${\cal O}(\as^n \log^n(s/|t|))$.

At NLL accuracy, the BFKL equation resums the corrections of ${\cal O}(\as^{n+1} \log^n(s/|t|))$~\cite{Fadin:1998py,Ciafaloni:1998gs,Kotikov:2000pm,Kotikov:2002ab}, by evaluating 
the radiative corrections to the gluon CEV. Namely,
its next-to-leading order (NLO) kernel is composed by the CEV for the emission of two gluons or a quark-antiquark pair~\cite{Fadin:1989kf,DelDuca:1995ki,Fadin:1996nw,DelDuca:1996nom} in next-to-multi-Regge kinematics (NMRK), where the two partons in the CEV are not ordered in rapidity (thus yielding, over the phase-space integration, a power of $\as$ but no powers of $\log(s/|t|)$), and by the one-loop corrections to the gluon CEV~\cite{Fadin:1993wh,Fadin:1994fj,Fadin:1996yv,DelDuca:1998cx,Bern:1998sc}. The ensuing soft divergences of the kernel are then regulated by the two-loop gluon Regge trajectory~\cite{Fadin:1995xg,Fadin:1996tb,Fadin:1995km,Blumlein:1998ib,DelDuca:2001gu}.

Underpinning the picture above is the fact that at LL and NLL accuracy, the virtual radiative corrections may be seen as corrections to the propagator of the gluon exchanged in the $t$ channel, a fact which is termed gluon Reggeization~\cite{Fadin:2006bj,Fadin:2015zea}, while the exchange of the gluon in the $t$ channel is called single Reggeon exchange. Beyond NLL accuracy, the single-Reggeon picture breaks down~\cite{DelDuca:2001gu}. Also (Regge) cut contributions occur, which may be interpreted as a
triple-Reggeon exchange~\cite{DelDuca:2001gu,DelDuca:2013ara,DelDuca:2014cya,Fadin:2016wso,Fadin:2017nka,Caron-Huot:2017zfo,Falcioni:2020lvv,Falcioni:2021buo}.
Therefore, the determination of the three-loop Regge trajectory~\cite{DelDuca:2021vjq,Falcioni:2021dgr,Caola:2021izf} and of the two-loop corrections to the gluon CEV~\cite{Buccioni:2024gzo,Abreu:2024xoh} require disentangling the single-Reggeon and triple-Reggeon contributions~\cite{Falcioni:2021dgr,Caola:2021izf,Buccioni:2024gzo,Abreu:2024mpk,Abreu:2024xoh}. 

After the disentangling of the single-Reggeon and triple-Reggeon contributions is done, one may consider carrying the BFKL program on to NNLL accuracy, by evaluating the next-to-next-to-leading order (NNLO) corrections to the kernel, which require the one-loop corrections to the CEV for the emission of two gluons or a quark-antiquark pair in NMRK -- so far, computed only for two gluons~\cite{Byrne:2022wzk} in 
${\cal N} = 4$ SYM -- and the CEV for the emission of three partons in next-to-next-to-multi-Regge kinematics (NNMRK)~\cite{DelDuca:1999iql,Antonov:2004hh,Duhr:2009uxa,Byrne:2025xxx}, in addition to the aforementioned two-loop corrections to the gluon CEV.

An amplitude, and thus a cross section, with exchange of a gluon ladder in the $t$ channel is then obtained by convoluting the ladder with process-dependent impact factors, which sit at the ends of the ladder. 
The accuracy in $\as$ at which impact factors are required is driven by the logarithmic accuracy of the gluon ladder: an amplitude for jet production at LL accuracy requires jet impact factors, and thus quark or gluon impact factors, at leading order in $\as$; for the same amplitude and for the jet cross section at NLL accuracy~\cite{Colferai:2010wu,Ducloue:2013hia}, jet impact factors at NLO in $\as$~\cite{Bartels:2001ge,Bartels:2002yj} are required. They are based on the one-loop impact factor~\cite{Fadin:1993wh,Fadin:1992zt,Fadin:1993qb,DelDuca:1998kx,Bern:1998sc}, and the impact factor for the emission of two gluons or of a quark-antiquark pair~\cite{Fadin:1989kf,DelDuca:1995ki,Fadin:1996nw,DelDuca:1996nom,DelDuca:1996km,Duhr:2009uxa}, evaluated in NMRK. Likewise, for amplitudes and for the jet cross sections at NNLL accuracy, jet impact factors at NNLO in $\as$ will be required. They will be built out of two-loop impact factor~\cite{DelDuca:2014cya,Caron-Huot:2017fxr,DelDuca:2021vjq,Falcioni:2021dgr,Caola:2021izf}, one-loop impact factors for the emission of two gluons or of a quark-antiquark pair, evaluated in NMRK~\cite{Canay:2021yvm,Byrne:2023nqx}, and the impact factors for the emission of three partons evaluated in NNMRK~\cite{DelDuca:1999iql}.

Likewise, the amplitude and the cross section for the production of a Higgs boson in association with a jet displays the exchange of a gluon ladder in the $t$ channel, which is convoluted with a jet impact factor and a Higgs impact factor at the ends of the ladder~\cite{Xiao:2018esv,Celiberto:2020tmb,Andersen:2022zte,Andersen:2023kuj}. The coupling of the Higgs boson to the gluons is mediated by a heavy-quark loop~\cite{Ellis:1975ap}, which in the Higgs Effective Field Theory (HEFT)~\cite{Wilczek:1977zn,Shifman:1978bx,Shifman:1979eb} may be replaced by an effective tree-level coupling. 
In Higgs boson production in association with a jet, the HEFT is a good approximation of the full theory as long as the jet transverse energies are smaller than the top-quark mass~\cite{Baur:1989cm}, $p_T < m_t$, and larger than the $b$-quark mass~\cite{Bonciani:2022jmb}, $p_T > m_b$, however the HEFT approximation is quite insensitive to the value of the Higgs–jet invariant mass~\cite{DelDuca:2001eu,DelDuca:2001fn}. 
As regards the inter-jet radiation in Higgs boson production in association with a jet, the BFKL resummation at LL accuracy yields results which are compatible with the full NLO computation~\cite{Andersen:2022zte}. This motivates evaluating $\as$ corrections to the BFKL ladder and thus to the Higgs impact factor, in order to improve the logarithmic accuracy.
At leading order in $\as$, the Higgs impact factor is known both in the full theory with heavy-quark mass dependence~\cite{DelDuca:2003ba} and in HEFT. At NLO, the Higgs impact factor is known only in HEFT~\cite{Hentschinski:2020tbi,Celiberto:2022fgx,Fucilla:2024cpf}. It is based on the one-loop Higgs impact factor, computed in HEFT~\cite{Hentschinski:2020tbi,Celiberto:2022fgx} and the impact factor for the emission of a Higgs and a gluon evaluated in NMRK, which is known both in the full theory with heavy-quark mass dependence~\cite{DelDuca:2003ba,Celiberto:2024bbv} and in HEFT~\footnote{In the MRK limit of this NMRK, also the leading-order Higgs CEV is known both in the full theory with heavy-quark mass dependence~\cite{DelDuca:2003ba} and in HEFT.}.

In this paper, we consider the HEFT two-loop amplitude for Higgs boson production in association with a jet, expanded to NNLL accuracy. Thanks to the simpler colour structure of the amplitudes for Higgs $+$ three partons with respect to parton-parton scattering, we are able to show that the Regge cuts do not contribute to the amplitudes through three loops. Thus, we determine the Higgs impact factor at two-loop accuracy. Accordingly, we are able to predict the single-logarithmic coefficient of the HEFT three-loop amplitude for Higgs boson production in association with a jet, expanded to NNLL accuracy.

In sec.~\ref{sec:hjampRegge}, we consider the amplitudes for Higgs $+$ three partons in the Regge limit, at tree-level in sec.~\ref{sec:hjtreeampRegge}, at NLL accuracy in sec.~\ref{sec:hjampReggenll} and at NNLL accuracy in sec.~\ref{sec:hjampReggennll}, where we discuss the issue of the Regge cut in this context.
In sec.~\ref{sec:amp2L}, we consider the two-loop amplitudes for Higgs $+$ three partons in general kinematics, expand them in the Regge limit and lay out their infrared structure.
Finally, in sec.~\ref{sec:IF2}, we present the Higgs impact factor at two-loop accuracy.
In sec.~\ref{sec:conc}, we draw our conclusions.
The paper is furnished with four appendices, which display the kinematics of Higgs $+$ three partons, the tree-level amplitudes in the spinor-helicity formalism, the anomalous dimensions which are used throughout the paper and provide the coefficients of the impact factors through two loops.

\section{The HEFT amplitudes for Higgs $+$ three partons in the Regge limit}
\label{sec:hjampRegge}

In the scattering between two partons of momenta $p_1$ and $p_2$,
with production of a Higgs boson of momentum $\ph$ with an associated jet, $p_1p_2\rightarrow p_3H$,
the relevant (squared) energy scales are the parton squared center-of-mass energy $s_{12}$, the momentum transfer $t=s_{13}$, 
the Higgs mass $\mh^2$ and the jet-Higgs invariant mass $\sah = (p_3 +\ph)^2$, where we identify the jet with the parton of momentum $p_3$.
Then the energy scales are related through momentum conservation,
\beq
s_{12} + s_{13} + s_{23} = \mh^2\, .\label{Hjmtmcons}
\eeq
In the Regge limit, app.~\ref{sec:mrk}, the light-cone momenta (\ref{in}) are strongly ordered (\ref{mrk}),
which entails that the rapidities are ordered as
\beq
\yh \gg y_3 + \left| \ln{\frac{\mht}{|p_{3_\perp}|}} \right|  \, ,\label{ymrk}
\eeq
where $\mht = \sqrt{|\pht|^2+\mh^2}$ is the transverse mass of the Higgs, with $|\pht|^2=p_{H,\,x}^2+p_{H\,,y}^2$ the Higgs momentum in the $(x,y)$ plane transverse to the beam axis $z$. We consider amplitudes for Higgs production in the Higgs Effective Field Theory (HEFT), app.~\ref{sec:hjamp},
where the loop-mediated Higgs-gluon coupling is replaced by an effective tree-level coupling
\begin{equation}
\label{eq:heftcoupl}
    {\mathcal{L}}_{\text{eff}} = -\frac{\lambda}{4}\,HG^{\mu\nu}_aG_{a;\mu\nu},
\end{equation}
where $H$ is the Higgs field and $G^{\mu\nu}_a$ is the gluon field strength. The Wilson coefficient $\lambda$, with the dimensions of the inverse of a mass, is written in term of the QCD coupling constant with $n_f$ light quarks and 1 heavy flavour. At the heavy quark scale, $m_t^2$, it reads \cite{Kniehl:1995tn,Spira:1995rr,Spira:1997dg,Chetyrkin:1997un,Schroder:2005hy,Chetyrkin:2005ia}
\begin{equation}
    \lambda=-\frac{\alpha_s^{(n_f+1)}(m_t^2)}{3\pi\,v}\left[1+11\,\left(\frac{\alpha_s^{(n_f+1)}(m_t^2)}{4\pi}\right)+{\cal{O}}(\alpha_s^2)\right],
\end{equation}
where $v$ is the vacuum expectation value of the Higgs. 
\subsection{The tree amplitudes for Higgs + three partons in the Regge limit}
\label{sec:hjtreeampRegge}

The tree amplitude for $g(p_1)\, g(p_2)\to g(p_3)\, H(\ph)$
can be written in the Regge limit as~\cite{DelDuca:2003ba}
\beq
\label{HgggatHE}
{\cal M}^{(0)}_{H3g}(p_1^{\nu_1}, p_2^{\nu_2}, p_3^{\nu_3}, \ph ) = 
\left[\frac{\lambda}{2}\, \delta^{a_2c} C^{{\sss H} (0)}(p_2^{\nu_2},\ph)\right] \frac{s}{t}
\left[\gs\, (F^{c})_{a_3a_1}\, C^{g(0)}(p_1^{\nu_1}, p_3^{\nu_3}) \right]\, ,
\eeq
with the incoming momenta parametrised as in \eqn{in}, and with $s=s_{12}=(p_1+p_2)^2$,
$q = p_1+p_3$,  $t=q^2 \simeq - |\qt|^2$ and $(F^{c})_{ab} = i \sqrt{2} f^{acb}$,
and where the superscript $\nu_i$ labels the helicity of gluon of momentum~$p_i$. 
We consider all the momenta as outgoing, so the helicities for incoming partons are reversed, see app.~\ref{sec:appa}. 
As it is apparent from the colour coefficient $(F^{a_3})_{a_1c} \delta^{a_2c}$, in \eqn{eq:hggg0amp} only the 
antisymmetric octet ${\bf 8}_a$ is exchanged in the $t$ channel. This remains true to all perturbative orders, as the amplitude is expected to be proportional to $(F^{a_3})_{a_1a_2}$. \eqn{HgggatHE} is written in terms of the
gluon impact factor, $g^*\, g \rightarrow g$, with $g^*$ an off-shell gluon~\cite{DelDuca:1995zy},
\begin{equation}
C^{g(0)}(p_1^\ominus, p_3^\oplus) = \frac{{p_{3\perp}^*}}{p_{3\perp}}\, ,
\label{eq:gif0}
\end{equation} 
with complex transverse coordinates $p_{\perp}$ as in \eqn{eq:scalprod}, and the Higgs impact factor, 
$g^*\, g \rightarrow H$~\cite{DelDuca:2003ba},
\begin{equation}
C^{{\sss H}(0)}(p_2^\oplus,\ph) = \qt\, .
\label{eq:hif0}
\end{equation} 
The impact factors (\ref{eq:gif0}) and (\ref{eq:hif0}) transform under parity into their complex conjugates,
\begin{equation}
[C^{g(0)}(p_1^\ominus, p_3^\oplus)]^\ast = C^{g(0)}(p_1^\oplus, p_3^\ominus)\,,\qquad 
[C^{\sss H(0)}(p_2^\oplus,\ph)]^\ast = C^{\sss H(0)}(p_2^\ominus,\ph) \,.
\label{eq:ifconj}
\end{equation} 
\eqn{HgggatHE} describes $2^3=8$ helicity configurations. However, at leading power in $t/s$, 
helicity is conserved on the tree-level gluon impact factor (\ref{eq:gif0}),
so in \eqn{HgggatHE} four helicity configurations are leading, two for each tree impact factor, \Eqns{eq:gif0}{eq:hif0}.
The helicity-flip impact factor $C^{g(0)}(p_1^\oplus, p_3^\oplus)$ and its parity conjugate $C^{g(0)}(p_1^\ominus,p_3^\ominus)$ are power suppressed in $t/s$. Multiplied by the Higgs impact factor $C^{{\sss H}(0)}(p_2^\oplus,\ph)$ and its parity conjugate $C^{{\sss H}(0)}(p_2^\ominus,\ph)$, they describe the
four helicity configurations which are power suppressed in $t/s$.

In the Regge limit, the amplitude for $q(\bar{q})\, g\to q(\bar{q})\, H$ scattering,
has the same analytic form as \eqn{HgggatHE}, up to
the replacement of an incoming gluon with a quark, or antiquark. 
For $q\, g\to q\, H$ scattering, that entails to replace in \eqn{HgggatHE} the gluon impact factor, where we set $\nu_3=-\nu_1$ in order to stress that helicity is conserved, with the quark impact factor,
\begin{equation}
\label{qlrag}
\left[\gs\, (F^{c})_{a_3a_1}\, C^{g(0)}(p_1^{\nu_1}, p_3^{-\nu_1}) \right] \leftrightarrow 
\left[\gs\, \sqrt{2}\, T^c_{i_3 \bar i_1}\, C^{q(0)}(p_1^{\nu_1}, p_3^{-\nu_1}) \right] \,,
\end{equation}
where
\begin{equation}
C^{q(0)}(p_1^\ominus, p_3^\oplus) = i\, \sqrt{ \frac{p_{3\perp}^*}{p_{3\perp}} }\, ,
\label{eq:qif0}
\end{equation}
which under parity transforms as
\begin{equation}
[C^{q(0)}(p_1^\ominus, p_3^\oplus)]^\ast = C^{q(0)}(p_1^\oplus, p_3^\ominus)\,.
\label{eq:ifqconj}
\end{equation} 
The appropriate replacement for the antiquark impact factor, required for $\bar{q}\, g\to \bar{q}\, H$ scattering, is 
\begin{equation}
\label{eq:qbarC}
    \left[\gs\, (F^{c})_{a_3a_1}\, C^{g(0)}(p_1^{\nu_1}, p_3^{-\nu_1}) \right] \leftrightarrow 
\left[-\gs\, \sqrt{2}\, T^c_{i_1 \bar i_3}\, C^{q(0)}(p_1^{\nu_1}, p_3^{-\nu_1}) \right],
\end{equation}
which differs from the quark impact factor in eq.~(\ref{qlrag}) by the generator of the group being in the conjugate representation, rather than in the fundamental.
In conclusion, in the high energy limit the amplitudes for Higgs+3 partons at tree level take the form
\begin{equation}
    {\cal{M}}_{i\, g\to i\, H}^{(0)}=\left[\frac{\lambda}{\sqrt{2}}\, \delta^{a_2c} C^{{\sss H} (0)}(p_2^{\nu_2},\ph)\right] \frac{s}{t}
\left[\gs\, (\mathbf{T}_i)^c_{a_3a_1}\, C^{i(0)}(p_1^{\nu_1}, p_3^{-\nu_1}) \right]\, ,
\end{equation}
where $\mathbf{T}_i$ is the colour generator in the representation of the parton $i$
\begin{align}
    ({\mathbf{T}}_i)^c_{ab}&=\left\{
    \begin{array}{rc}
    -i\,f^{abc},     &  i = g,\\
    T^{c}_{ab},      &  i = q,\\ 
    -T^{c}_{ba},     &  i = \bar{q}.
    \end{array}
    \right.
\end{align}

\subsection{The amplitudes for Higgs + three partons at NLL accuracy}
\label{sec:hjampReggenll}

Since only the antisymmetric octet ${\bf 8}_a$ is exchanged in the $t$ channel,
which is odd under $s\leftrightarrow u$ crossing, we expect that also the kinematic part
of the amplitudes for Higgs $+$ three partons is odd under $s\leftrightarrow u$ crossing.
Then at next-to-leading-logarithmic (NLL) accuracy, we write the amplitude for
Higgs $+$ three gluons as~\cite{Fadin:1993wh}
\beqa
\label{Hggnll}
\lefteqn{ {\cal M}_{H3g}(p_1^{\nu_1}, p_2^{\nu_2}, p_3^{\nu_3}, \ph ) } \\
&=& \left[\frac{\lambda}{2}\, \delta^{a_2c} C^{{\sss H}}(p_2^{\nu_2},\ph)\right] 
\frac{s}{2t} \left[ \left(\frac{s}{\tau}\right)^{\alpha(t)} + \left(\frac{-s}{\tau}\right)^{\alpha(t)} \right]
\left[\gs\, (F^{a_3})_{a_1c}\, C^{g}(p_1^{\nu_1}, p_3^{-\nu_1}) \right]\, , \nn
\eeqa
where $\tau > 0$ is a Regge factorisation scale, which is of order of $t$, and much smaller than $s$.
In \eqn{Hggnll}, $\alpha(t)$ is the gluon Regge trajectory, whose expansion in $\as$ is
\begin{equation}
\alpha(t) = \frac{\as}{4\pi} \alpha^{(1)}(t) + 
\left(\frac{\as}{4\pi}\right)^2 \alpha^{(2)}(t) + \ord(\as^3)\,
,\label{alphb}
\end{equation}
with (unrenormalised) coefficients \cite{Lipatov:1976zz,Kuraev:1976ge,Fadin:1995xg,Fadin:1996tb,Fadin:1995km,Blumlein:1998ib,DelDuca:2001gu},
\beqa
\alpha^{(1)}_{\text{bare}}(t) &=& C_A\, \frac{2}{\epsilon} 
\left(\frac{\mu^2}{-t}\right)^{\epsilon} \kappa_{\Gamma}\, ,\label{alph} \\
\alpha^{(2)}_{\text{bare}}(t) &=& \kappa_{\Gamma}^2 \left(\frac{\mu^2}{-t}\right)^{2\epsilon} 
\left[ \frac{\beta_0\gamma_K^{(1)}}{2\eps^2} C_A + \frac{2\gamma_K^{(2)}}{\eps} C_A 
+ \left(\frac{404}{27} - 2\zeta_3\right) C_A^2 - \frac{56}{27}\, C_An_f \right.\nonumber\\
&+&\epsilon\left(C_A\left(\frac{2428}{81}-66\zeta_3-8\zeta_4\right)+n_f\left(-\frac{328}{81}+12\zeta_3\right)\right)+\epsilon^2\left(C_A\left(\frac{14576}{243}-134\zeta_3\right.\right.\nonumber\\
&-&\left. \left. \left. 99\zeta_4+36\zeta_2\zeta_3+82\zeta_5\right)+n_f\left(-\frac{1952}{243}+20\zeta_3+18\zeta_4\right)\right)\right] \,,\label{alph2}
\eeqa
through $\ord(\eps^2)$, with $C_A=N_c$ the number of colours, $n_f$ the number of light quark flavours,
$\beta_0$ the one-loop coefficient of the beta function and $\gamma_K^{(2)}$ the two-loop coefficient of the cusp anomalous dimension, app.~\ref{AppAnDim}, and 
\begin{equation}
\kappa_\Gamma = (4\pi)^\epsilon\, {\Gamma(1+\epsilon)\,
\Gamma^2(1-\epsilon)\over \Gamma(1-2\epsilon)}\, .\label{cgam}
\end{equation}
The coefficients of the renormalised Regge trajectory in $\overline{\text{MS}}$ scheme read
\begin{align}
\label{eq:agNLLren}
    \alpha^{(1)}(t,\mu^2) &= \alpha^{(1)}_{\text{bare}}(t),\\
    \alpha^{(2)}(t,\mu^2) &= \alpha^{(2)}_{\text{bare}}(t)-\frac{\beta_0}{\epsilon}\alpha^{(1)}_{\text{bare}}(t).
\end{align}
In \eqn{Hggnll}, the gluon impact factor $C^{g}$ is expanded in $\as$ as
\begin{equation}
C^{g}(p_1^{\nu_1};p_3^{-\nu_1}) = 
C^{g (0)}(p_1^{\nu_1};p_3^{-\nu_1}) \left(1 + \frac{\as}{4\pi} c^{g (1)}(t,\tau,\mu^2) + \ord(\as^2) \right)\, 
.\label{fullv}
\end{equation}
where $C^{g (0)}$ is given in \eqn{eq:gif0} and
the one-loop coefficient $c^{g (1)}$ is real and independent of the helicity configuration. Its unrenormalised version through $\ord(\eps^0)$ is~\cite{Fadin:1993wh,Fadin:1992zt,Fadin:1993qb,DelDuca:1998kx,Bern:1998sc,DelDuca:2017pmn}
\beqa
\label{impactcorrect}
\lefteqn{ c^{g (1)}_{\text{bare}}(t,\tau) } \\ &=&  \kappa_\Gamma \left({\mu^2\over -t}\right)^{\epsilon} \left[
- \frac{\gamma_K^{(1)} C_A}{\epsilon^2} + \frac{4\gamma_g^{(1)} }{\epsilon} + \frac{\beta_0}{2\epsilon} 
+ \frac{C_A}{\epsilon} \log\left(\frac{\tau}{-t}\right)  
- \gamma_K^{(2)} + 2\zeta_2C_A  \right] \nn\\
&=&  \kappa_\Gamma  \left( \frac{\mu^2}{-t} \right) ^\epsilon 
\left[ \left( -\frac{2}{\epsilon^2} -\frac{11}{6\epsilon} + \frac1{\epsilon} \log\left(\frac{\tau}{-t}\right)
- \frac{32}{9} - \frac{\delta_R}{6} + \frac{\pi^2}{2} \right) N_c
+ \left( \frac1{3\epsilon} + \frac{5}{9} \right) n_f \right] \,, \nn
\eeqa
where $\delta_R$ is a regularisation parameter, which labels the computation as done in conventional dimensional 
regularization (CDR)/'t-Hooft-Veltman (HV) schemes, for 
$\delta_R =1$~\cite{Fadin:1993wh,Fadin:1992zt,DelDuca:1998kx,Bern:1998sc}, or
in the dimensional reduction (DR)/ four dimensional helicity (FDH) schemes, for $\delta_R =0$~\cite{DelDuca:1998kx,Bern:1998sc}.
In \eqn{impactcorrect}, the infrared $\epsilon$ poles are accounted for by the cusp anomalous dimension
and by the gluon collinear anomalous dimension~\cite{DelDuca:2014cya},
with $\gamma_K^{(1)}$ the one-loop coefficient of the cusp anomalous dimension (\ref{eq:k2}), 
$\gamma_g^{(1)}$ the one-loop coefficient of the gluon collinear anomalous dimension (\ref{eq:c1}). Note that the one-loop coefficient, $c^{g (1)}$, is known in the HV scheme to all orders in $\epsilon$~\cite{Fadin:1993wh,Fadin:1992zt,Bern:1998sc}. The $\overline{\rm MS}$-renormalised one-loop coefficient is 
\begin{equation}
\label{impactcorrectREN}
    c^{g (1)}(t,\tau) = c^{g (1)}_{\text{bare}}(t,\tau) - \frac{\beta_0}{2\epsilon}.
\end{equation}
Finally, eq.~(\ref{Hggnll}) involves the impact factor of the Higgs boson
\begin{equation}
\label{eq:expaCH}
    C^{{\sss H}}(p_2^{\nu_2},\ph) = C^{H (0)}(p_2^\nu,\ph)\,\left(1 + \frac{\alpha_s}{4\pi} c^{H(1)}(t,m_H^2,\tau) + {\cal{O}}(\alpha_s^2)\right),
\end{equation}
where $C^{H (0)}(p_2^\nu,\ph)$ is given in eq.~(\ref{eq:hif0}). The one-loop coefficient has been computed up to ${\cal{O}}(\epsilon^0)$ ~\cite{Celiberto:2022fgx,Fucilla:2024cpf} and its unrenormalised value is
\begin{align}
    \begin{split}
    \label{eq:cH1}
        c^{H (1)}_{\text{bare}}\left(t,m_H^2,\tau\right) &=\kappa_\Gamma\left(\frac{\mu^2}{-t}\right)^\epsilon\left[-\frac{\gamma_K^{(1)}C_A}{2\epsilon^2}+\frac{2\gamma_g^{(1)}}{\epsilon}+\frac{\beta_0}{\epsilon}+\frac{C_A}{\epsilon}\log\left(\frac{\tau}{-t}\right)+C_A\left(2\text{Li}_2\left(\frac{t}{m_H^2}\right)\right.\right.\\
        &\hspace{-2cm}\left.\left.+2\log\left(\frac{-t}{m_H^2}\right)\log\left(\frac{m_{H\perp}^2}{m_H^2}\right)-\log^2\left(\frac{-t}{m_H^2}\right)+\frac{67}{18}+5\zeta_2+2i\pi\log\left(\frac{m_{H\perp}^2}{-t}\right)\right)-\frac{5}{9}n_f\right],\\
        &\hspace{-2cm}=\kappa_\Gamma\left(\frac{\mu^2}{-t}\right)^\epsilon\left[\left(-\frac{1}{\epsilon^2}+\frac{11}{6\epsilon}+\frac{1}{\epsilon}\log\left(\frac{\tau}{-t}\right)+2\text{Li}_2\left(\frac{t}{m_H^2}\right)+2\log\left(\frac{-t}{m_H^2}\right)\log\left(\frac{m_{H\perp}^2}{m_H^2}\right)\right.\right.\\
        &\hspace{-2cm}\left.\left.-\log^2\left(\frac{-t}{m_H^2}\right)+\frac{67}{18}+5\zeta_2+2i\pi\log\left(\frac{m_{H\perp}^2}{-t}\right)\right)N_c-\left(\frac{1}{3\epsilon}+\frac{5}{9}\right)n_f\right].
    \end{split}
\end{align}
The renormalised coefficient in the $\overline{\rm MS}$ scheme is 
\begin{equation}
    c^{H(1)}(t,m_H^2,\tau) = c^{H(1)}_{\text{bare}}(t,m_H^2,\tau) - \frac{\beta_0}{\epsilon}.
\end{equation}

At NLL accuracy, the amplitude for $q(\bar{q})+\, g\to q(\bar{q})+\, H$ scattering can be obtained from \eqn{Hggnll} by choosing the colour generator in the appropriate representation and replacing
the gluon impact factor with the quark impact factor, as described in eqs.~(\ref{qlrag}) and (\ref{eq:qbarC}). The quark impact factor is expanded in $\as$ as in \eqn{fullv}, with
$C^{q (0)}$ as in \eqn{eq:qif0} and with (unrenormalised) one-loop coefficient, which has been computed for 
$\delta_R =1$~\cite{DelDuca:1998kx,Fadin:1993qb} and for $\delta_R =0$~\cite{DelDuca:1998kx},
\beqa
\lefteqn{ c^{q (1)}_{\text{bare}}(t,\tau) } \\ &=&  \kappa_\Gamma \left({\mu^2\over -t}\right)^{\epsilon} \left[
- \frac{\gamma_K^{(1)} C_F}{\epsilon^2} + \frac{4\gamma_q^{(1)} }{\epsilon} + \frac{\beta_0}{2\epsilon} 
+ \frac{C_A}{\epsilon} \log\left(\frac{\tau}{-t}\right)  + \gamma_K^{(2)} + \left(1+4\zeta_2\right) C_A  
- (7 + \delta_R) C_F \right] \nn\\
&=&  \kappa_\Gamma  \left( \frac{\mu^2}{-t} \right) ^\epsilon 
\left[ \left( -\frac1{\epsilon^2} +\frac1{3\epsilon} + \frac1{\epsilon} \log\left(\frac{\tau}{-t}\right)
+ \frac{19}{18} - \frac{\delta_R}3 + \frac{\pi^2}2 \right) N_c - \left( \frac1{3\eps} + \frac{5}{9} \right) N_f 
\right. \nn\\ && \left. \qquad\qquad 
+ \left( \frac1{\epsilon^2} +\frac3{2\epsilon} + \frac{7+\delta_R}2 \right) \frac1{N_c} \right] \,, \nn
\label{eq:qif1}
\eeqa
and the corresponding $\overline{\text{MS}}$-renormalised value is
\begin{equation}
    c^{q(1)}(t,\tau,\mu^2) = c^{q (1)}_{\text{bare}}(t,\tau) - \frac{\beta_0}{2\epsilon}.
\end{equation}

\subsection{The Regge pole and the Regge cuts in the amplitudes at NNLL}
\label{sec:hjampReggennll}

The factorised expression of eq.~(\ref{Hggnll}) is dictated by the exchange of a reggeized gluon in the $t$-channel \cite{Lipatov:1976zz,Grisaru:1973ku,Grisaru:1973vw,Grisaru:1973wbb,Fadin:1975cb,Fadin:1995xg,Fadin:2006bj,Ioffe:2010zz} and is expected to break down at NNLL accuracy, due to the contribution of \textit{Regge cuts} \cite{DelDuca:2001gu}. In general, the amplitudes take the form
\beqa
\label{Hggnnll}
\lefteqn{ {\cal M}_{i\,g\to i\,H}(p_1^{\nu_1}, p_2^{\nu_2}, p_3^{\nu_3}, \ph ) } \\
&=& \left[\frac{\lambda}{\sqrt{2}}\, \delta^{a_2c} C^{{\sss H}}(p_2^{\nu_2},\ph)\right] 
\frac{s}{2t} \left[ \left({s\over \tau}\right)^{\alpha(t)} + \left({-s\over \tau}\right)^{\alpha(t)} \right]
\left[\gs\, (\mathbf{T}_i^{c})_{a_3a_1}\, C^{i}(p_1^{\nu_1}, p_3^{-\nu_1}) \right]\,+ {\cal{M}}_{\text{cut},\,i}. \nn
\eeqa
At NNLL the equation above requires the terms of ${\cal{O}}(\alpha_s^2)$ in the expansion of the impact factors, eqs.~(\ref{fullv}) and (\ref{eq:expaCH}), as well as the coefficients up to ${\cal{O}}(\alpha_s^3)$ of the gluon Regge trajectory, eq.~(\ref{alphb}). 
In order to make use of eq.~(\ref{Hggnnll}), we need a prescription \cite{DelDuca:2013ara,DelDuca:2014cya,Caron-Huot:2017fxr,Fadin:2016wso,Fadin:2017nka,Falcioni:2021buo,Falcioni:2021dgr} to disentangle the contribution of the reggeized gluon from the Regge cut 
\begin{equation}
{\cal{M}}_{\text{cut},\,i} = \sum_{n\geq 2}\left(\frac{\alpha_s}{4\pi}\right)^n{\cal{M}}_{\text{cut},\,i}^{(n)}.
\end{equation}
Here we follow~\cite{Falcioni:2021buo}, where ${\cal{M}}^{(n)}_{\text{cut}}$ has been defined through NNLL accuracy for the $2\to2$ amplitudes of 4 coloured partons. By taking one parton as a colour singlet, we have
\begin{align}
\label{eq:cutL}
    {\cal{M}}_{\text{cut},\,i}^{(n)}&=\left[\frac{3-n}{(n-2)!}(\alpha^{(1)}(t))^{n-2}\hat{M}_{\text{cut},\,i}^{(2)}+\frac{(\alpha^{(1)}(t))^{n-3}}{(n-3)!}\hat{M}_{\text{cut},\,i}^{(3)}+\sum_{m=0}^{n-4}\frac{(\alpha^{(1)}(t))^m}{m!}\hat{M}_{\text{cut},\,i}^{(n-m)}\right]\nn\\
    &\times\left(\log\frac{s}{-t}-i\frac{\pi}{2}\right)^{n-2},
\end{align}
where we use the signature-even logarithm \cite{Caron-Huot:2013fea}
\begin{equation}
\label{eq:SymLog2to2}
    \frac{1}{2}\left(\log\frac{u}{t} + \log\frac{s}{t}\right) = \log\frac{s}{-t} -\frac{i\pi}{2} + {\cal{O}}\left(\frac{t}{s}\right).
\end{equation}
The {\textit{genuine}} $n$-loop contributions to the cut, $\hat{M}^{(n)}_{\text{cut}}$, are given by the colour rotations of the tree-level amplitude, originated by the exchange of multi-Reggeon states between the target and the projectile particles~\cite{Caron-Huot:2013fea,Caron-Huot:2017fxr}. In the prescription of ref.~\cite{Falcioni:2021buo}, ${\hat{M}^{(n)}_{\text{cut}}}$ is {\textit{defined}} to include only subleading contributions in the large-$N_c$ limit. In the case of the amplitudes for Higgs + 3 partons, such prescription implies that the Regge cut cannot contribute up to four loops, as shown below.
 
It is convenient to write the colour rotations in terms of the generators in the $s$-, $t$- and $u$-channel~\cite{Dokshitzer:2005ig}
\beqa
  {\bf T_s} & \equiv & {\bf T}_1 + {\bf T}_2 \, , \nonumber \\
  {\bf T_t} & \equiv & {\bf T}_1 + {\bf T}_3 \, , \nonumber \\
  {\bf T_t} & \equiv & {\bf T}_2 + {\bf T}_3 \, ,
\label{Tstudef}
\eeqa
where the colour-insertion operator ${\bf T}_i$~\cite{Bassetto:1983mvz,Catani:1996vz} acts as the identity on the colour indices of all external partons other than parton $i$, and 
it inserts a colour generator in the appropriate representation on
the $i$-th leg.  In addition, one may
define the product ${\bf T}_i \cdot {\bf T}_j \equiv \sum_a 
{\bf T}_i^a \, {\bf T}_j^a$, where $a$ is the adjoint index 
counting over the color generators. Then ${\bf T}_i^2
\equiv {\bf T}_i \cdot {\bf T}_i = C_i$, where $C_i$ is the quadratic 
Casimir eigenvalue appropriate for the color representation of parton $i$.
Since the amplitudes, eq.~(\ref{Hggnnll}), have {\textit{odd signature}}, the most general colour rotation induced by multi-Reggeon exchanges, $\hat{M}^{(n)}_{\text{MRS},\,i}$, has the structure \cite{Caron-Huot:2017fxr,Falcioni:2020lvv}
\begin{align}
\label{def:tchanQuadCol}
    \hat{M}^{(n)}_{\text{MRS},\,i} &= P^{(n)}(\mathbf{T_t^2}, \mathbf{T_{s-u}^2})\,\mathcal{M}^{(0)}_{i\,g\to i\,H},  
\end{align}
where $\mathbf{T^2_{s-u}}=\frac{1}{2}(\mathbf{T_s^2}-\mathbf{T_u^2})$. $P^{(n)}$ is a polynomial in the colour operators $\mathbf{T^2_t}$ and $\mathbf{T^2_{s-u}}$, featuring an even number of powers of $\mathbf{T^2_{s-u}}$. The coefficients of $P^{(n)}$ depend on the dimensional regulator $\epsilon$ and, in principle, on Casimir invariants in the adjoint representation\footnote{The diagrams that enter $\hat{M}_{\text{MRS},\,i}$ are purely {\textit{gluonic}} \cite{Falcioni:2020lvv}, thus universal in every gauge theory.} \cite{Falcioni:2021buo}. Using colour conservation for the three-parton scattering
\beq
  ({\bf T}_1 + {\bf T}_2 + {\bf T}_3)\mathcal{M}_{i\,g\to i\,H} = 0 \, ,
\label{color_conservation} 
\eeq
where $\mathcal{M}_{i\,g\to i\,H}$ indicates every (colour-singlet) amplitude, we get 
\begin{align}
    &\mathbf{T^2_t} = C_A\,\mathbf{1} = N_c\,\mathbf{1},    &\mathbf{T^2_{s-u}} = 0,
\label{def:Tt2Tsu0}
\end{align}
for both the $q+g\to q+H$ and the $g +g\to g+H$ scattering. For $n\leq3$, the coefficients in $P^{(n)}$ can only contain powers of the quadratic Casimir $C_A$ and  eq.~(\ref{def:tchanQuadCol}) takes the form
\begin{equation}
\label{eq:MRS}
    \hat{M}^{(n)}_{\text{MRS},\,i} = f_{n}(\epsilon)\,C_A^n\,\mathcal{M}^{(0)}_{i+g\to i+H},
\end{equation}
which is the only term that we can construct up to three-loop order\footnote{Even at higher loop order, the structure of $\hat{M}_{\text{MRS},\,i}$ is tightly constraint by colour conservation, with
\begin{equation*}
    P^{(n)}(\mathbf{T^2_t},\mathbf{T^2_{s-u}}) = \sum_{k=0}^n c_k(C_A,d_{AA},\dots)\,(\mathbf{T^2_t})^k =\sum_{k=0}^n c_k(C_A,d_{AA},\dots)\,C_A^k,
\end{equation*}
where $c_k$ is a polynomial in the Casimir invariants of the adjoint representation, with $d_{AA}$ defined in eq.~(\ref{def:dAA}), since $P^{(n)}$ arises from gluonic diagrams \cite{Falcioni:2020lvv}. The monomials of $c_k$ are such that each term in $P^{(n)}$ is a $n$-loop colour factor. Therefore, we conjecture
\begin{align*}
    P^{(4)} &= \tilde{c}_1(\epsilon) C_A^4 + \tilde{c}_2(\epsilon) \frac{d_{AA}}{N_A},\\
    P^{(5)} &= \tilde{c}_3(\epsilon) C_A^5 + \tilde{c}_4(\epsilon) C_A\frac{d_{AA}}{N_A}.
\end{align*}
}, using the available Casimir invariant $C_A$ and the identity in colour space. Eq.~(\ref{eq:MRS}) has no subleading contribution in the large $N_c$ limit, hence we define
\begin{equation}
\label{eq:cut2;;3}
    \hat{M}^{(2)}_{\text{cut},\,i} = \hat{M}^{(3)}_{\text{cut},\,i} = 0.
\end{equation}

The scheme choice in eq.~(\ref{eq:cut2;;3}) implies that only the Regge pole contributes to the amplitude through three loops. By expanding eq.~(\ref{Hggnnll}) in $\alpha_s/(4\pi)$ up to ${\cal{O}}(\alpha_s^2)$ we get
\begin{align}
  \begin{split}
  \label{eq:Hgggnnll2}
      {\cal{M}}^{(2)}_{i\,g\to i\,H}&= \frac{\left(\alpha^{(1)}\right)^2}{2}\,L^2+\left[\alpha^{(2)}+\alpha^{(1)}(c^{i(1)}+c^{H(1)})\right]\,L+c^{i(2)}+c^{H(2)}\\ 
      &+c^{\text{i}(1)}\,c^{H(1)}-\frac{\pi^2}{8}\left(\alpha^{(1)}\right)^2,
  \end{split}  
\end{align}
where we used the notation ${\cal{M}}_{ig\to iH}=\sum {\cal{M}}^{(n)}_{ig\to iH}(\alpha_s/(4\pi))^n$ and $L=\log\frac{s}{\tau}-i\frac{\pi}{2}$. We omit the arguments of $\alpha^{(p)}$, $c^{H(p)}$ and $c^{i(p)}$, for $p=1,2$, to simplify the notation. The coefficients $\alpha^{(1)}$ and $\alpha^{(2)}$ are given in eq.~(\ref{alph}) and (\ref{alph2}), respectively. Note that the equation above requires the one-loop impact factors, $c^{i(1)}$ and $c^{H(1)}$, to ${\cal{O}}(\epsilon^2)$. These are obtained from eqs.~(\ref{eq:IF1loopGEN}) and (\ref{eq:IFHiggs1GEN}), respectively. In addition, the two-loop impact factors of the quark and of the gluon $c^{i(2)}$, for $i= q,g$, which were computed in~\cite{DelDuca:2014cya,Caron-Huot:2017fxr,DelDuca:2021vjq,Falcioni:2021dgr,Caola:2021izf}, are given by plugging eqs.~(\ref{eq:IF2loopG}) and (\ref{eq:IF2loopQ}) into eq.~(\ref{eq:IF2loopGEN}). Therefore, eq.~(\ref{eq:cut2;;3}) determines the amplitudes for both $q+g\to q+H$ and $g+g\to g+H$ in terms of the two-loop Higgs impact factor, $c^{H(2)}$, as we will discuss in sec.~\ref{sec:IF2}. By proceeding to three loops we get
\begin{align}
    \begin{split}
    \label{eq:Hgggnnll3}
        {\cal{M}}^{(3)}_{i\,g\to i\,H}&=\frac{\left(\alpha^{(1)}L\right)^3}{6}+\alpha^{(1)}L^2\left[\frac{\alpha^{(1)}}{2}\left(c^{i(1)}+c^{H(1)}\right)+\alpha^{(2)}\right]+L\left[\vphantom{\frac{1}{2}}\alpha^{(3)}+\alpha^{(2)}\left(c^{i(1)}+c^{H(1)}\right)\right.\\
        &\left.+\alpha^{(1)}\left(c^{i(2)}+c^{H(2)}+c^{i(1)}c^{H(1)}-\frac{\pi^2}{8}\left(\alpha^{(1)}\right)^2\right)\right]  + {\cal{O}}(L^0),
    \end{split}
\end{align}
where the three-loop Regge trajectory~\cite{DelDuca:2021vjq,Falcioni:2021dgr,Caola:2021izf} is written as
\begin{align}
    \alpha^{(3)}(t,\mu^2)&=\alpha^{(3)}_{\text{bare}}(t)+\alpha^{(1)}_{\text{bare}}(t)\frac{\beta_0^2}{\epsilon^2}-\frac{1}{2\epsilon}\left(\alpha^{(1)}_\text{bare}(t)\beta_1+4\alpha^{(2)}_{\text{bare}}(t)\beta_0\right),
\end{align}
where $\alpha^{(1)}_{\text{bare}}$ and $\alpha^{(2)}_{\text{bare}}$ are given in eqs.~(\ref{alph}) and (\ref{alph2}), respectively, and
\begin{align}
   \alpha^{(3)}_{\text{bare}}(t)&=\kappa_\Gamma^3\left(\frac{\mu^2}{-t}\right)^{3\epsilon}C_A\Bigg[\frac{\beta_0^2\gamma_K^{(1)}}{3\epsilon^3}+\frac{\beta_1\gamma_K^{(1)}+16\beta_0\gamma_K^{(2)}}{6\epsilon^2}+\frac{1}{\epsilon}\left(\frac{4\beta_0}{27}(C_A(202-27\zeta_3)-28n_f)\right.\nonumber\\
   &\left.+\frac{16}{3}\gamma_K^{(3)}\right)+C_A^2\left(\frac{617525}{1458}+\frac{3196}{81}\zeta_2-\frac{19732}{27}\zeta_3-\frac{253}{3}\zeta_4+\frac{40}{3}\zeta_2\zeta_3+16\zeta_5\right)\nonumber\\
   &+C_An_f\left(-\frac{82097}{729}+\frac{412}{81}\zeta_2+\frac{2140}{9}\zeta_3+\frac{22}{3}\zeta_4\right)+C_Fn_f\left(-\frac{1711}{54}+\frac{152}{9}\zeta_3\right.\nonumber\\
   &+8\zeta_4\bigg)+n_f^2\left(\frac{4864}{729}-\frac{560}{27}\zeta_3\right)\Bigg]. 
\end{align}
Hence, once we fix the expression for $c^{H(2)}$, we obtain a prediction for the three-loop amplitudes through ${\cal{O}}(L)$ in the high-energy limit.

Finally, starting at four-loop order, the coefficients of the polynomial $P^{(n)}$ in eq.~(\ref{def:tchanQuadCol}) depend also on quartic Casimir invariants, which involve subleading terms in the large-$N_c$ expansion. Therefore, we might have non-planar contributions \cite{Falcioni:2020lvv} $\hat{M}^{(4)}_{\text{cut},\,i}\propto\frac{d_{AA}}{N_A}-\frac{C_A^4}{24}$, where $N_A$ is the dimension of the adjoint representation and
\begin{equation}
\label{def:dAA}
    d_{AA}=\frac{1}{4!}\sum_{\sigma\in{\cal{S}}_4}\text{Tr}\left[F^{\sigma(a)}F^{\sigma(b)}F^{\sigma(c)}F^{{\sigma(d)}}\right]\,\text{Tr}\left[F^aF^bF^cF^d\right].
\end{equation}
For the $\text{SU}(N_c)$ gauge group we obtain $\frac{d_{AA}}{N_A}=N_c^2(N_c^2+36)/24$, such that $\frac{d_{AA}}{N_A}-\frac{C_A^4}{24}$ is indeed suppressed for large $N_c$. To determine whether $\hat{M}^{(4)}_{\text{cut},\,i}$ vanishes we must perform a genuine calculation of the multi-Regge diagrams by extending the formalism \cite{Caron-Huot:2013fea} to the case of one colourless off-shell parton\footnote{It would be wrong to naively impose $\mathbf{T^2_{s-u}}=0$ onto the result for $\hat{M}_{\text{cut}}^{(4)}$ given in \cite{Falcioni:2021buo}, because it would immediately violate the infrared structure of the four-loop amplitudes~\cite{Gardi:2009qi,Becher:2019avh}. Indeed, the derivation in ref.~\cite{Falcioni:2021buo} relies explicitly on colour conservation of four partons.}, which we leave to future work.

\section{Comparison with explicit results in general kinematics}
\label{sec:amp2L}

In this section we take the limit of the amplitudes for Higgs + 3 partons through two loops and we compare them with the asymptotic behaviour described in sec.~\ref{sec:hjampRegge}, in order to verify the conditions in eq.~(\ref{eq:Hgggnnll2}) and (\ref{eq:Hgggnnll3}). 
The amplitudes for $g\,g\to g\,H$ and $q\,g\to q\,H$ have been computed up to the terms of ${\cal{O}}(\epsilon^0)$ at one loop in ref.~\cite{Schmidt:1997wr} and at two loops in refs.~\cite{Gehrmann:2011aa,Duhr:2012fh}. More recently, the one-loop amplitudes have been computed through ${\cal{O}}(\epsilon^4)$ and the two-loop amplitudes up to ${\cal{O}}(\epsilon^2)$ \cite{Gehrmann:2023etk}.
Here we utilize the latter results, which are given in the decay region for 
\begin{align}
\label{def:gluDecay}
    H \to g_{+}(p_1) + g_{+}(p_2) +g_{-}(p_3),\\
\label{def:quaDecay}    
    H \to q_L(p_1) + \bar{q}_R(p_2) + g_{+}(p_3).
\end{align}
The analytic continuation to the physical region of the scattering processes
\begin{align}
\label{def:gluScat}
g_{-}(p_1) + g_{+}(p_3) \to H + g_{+}(p_2),\\
\label{def:quaScat}
q_{R}(p_1) + g_{-}(p_3) \to H + q_{R}(p_2),
\end{align}
is described in refs.~\cite{Gehrmann:2023etk,Gehrmann:2002zr}. After this step, the amplitudes are expressed in terms of harmonic polylogarithms (HPLs) \cite{Remiddi:1999ew} in the variable $v=\frac{m_H^2}{s_{13}}\in[0,1]$ and two-dimensional harmonic polylogarithms (2dHPLs) \cite{Gehrmann:2000zt,Gehrmann:2001ck} with indices in the alphabet $\{0,1,-v,1-v\}$ and argument $u=-\frac{s_{23}}{s_{13}}\in[0,1-v]$. Note that the notation of ref.~\cite{Gehrmann:2023etk}, which is kept in eqs.~(\ref{def:gluDecay})-(\ref{def:quaScat}), differ from the convention of this paper. We write the amplitudes in terms of the variables introduced in eq.~(\ref{HgggatHE}) by replacing
\begin{align}
    v&=\frac{m_H^2}{s},\\
    u&=\left\{
    \begin{array}{lll}
         -\frac{t}{s}&  &\text{gluon scattering, eq.~(\ref{def:gluScat})}\\
         1-\frac{m_H^2-t}{s}& &\text{quark scattering, eq.~(\ref{def:quaScat})}
    \end{array}
    \right.
\end{align}
\subsection{Expansion of the amplitudes in the Regge limit}
\label{sec:M2Regge}
In order to derive the high-energy limit of the amplitudes, we perform the asymptotic expansion of the HPLs and of the 2dHPLs. The former ones are expanded around $v\to 0$ using \texttt{PolyLogTools} \cite{Duhr:2019tlz}. Regarding the 2dHPLS, their argument $u$ approaches $u\to0$ for gluon-gluon scattering, eq.~(\ref{def:gluScat}), and $u\to1$ for quark-gluon scattering, eq.~(\ref{def:quaScat}). In each amplitude, we select a set of 2dHPLs of transcendental weight $w\leq6$ that are independent under shuffle relations. We denote either of them as
\begin{equation}
    \vec{f}(v,x) = \left(
    \begin{array}{c}
    \varepsilon^6\,G(0,0,0,0,0,1-v,u)\\
    \cdots\\
    \varepsilon^w\,G(a_1,\dots\,a_w,u)\\
    \cdots\\
    1
    \end{array}
    \right),
\end{equation}
where the parameter $\varepsilon$ keeps track of the transcendental weight and $x=-t/m_H^2$. The 2dHPLs obey linear differential equations\footnote{The set of functions in $\vec{f}$ might be suitably extended to include 2dHPLs that do not appear in the amplitudes, but do arise in the derivatives with respect to $v$ and $x$.}
\begin{align}
\label{deq:v}
    \frac{\partial}{\partial v}\vec{f}(v,x) &= \varepsilon\,\mathbf{M_v}(v,x) \,\vec{f}(v,x),\\
\label{deq:x}    
    \frac{\partial}{\partial x}\vec{f}(v,x) &= \varepsilon\,\mathbf{M_x}(v,x) \,\vec{f}(v,x),    
\end{align}
where we obtained the matrices $\mathbf{M_v}$ and $\mathbf{M_x}$ by computing the derivatives of the 2dHPL with \texttt{PolyLogTools}. The boundary conditions were computed by evaluating the 2dHPL at the point $(v=1/4,x=1)$ using \texttt{GiNaC} \cite{Vollinga:2004sn}. We solve the systems above in a generalised series expansion around $v\to0$ with
\begin{equation}
\label{def:MvEXP}
    \mathbf{M}_v = \frac{1}{v}\,\mathbf{M_v^{(0)}} + \sum_{k\geq0} v^k\,\mathbf{M_v^{(1+k)}}(x),. 
\end{equation}
where $\mathbf{M_v^{(0)}}$ is a matrix of rational numbers. We use the package \texttt{DiffExp} \cite{Hidding:2020ytt} to transport the boundary conditions to $v\to 0$, where they diverge logarithmically  
\begin{align}
    \lim_{v\to0} \vec{f}(v,1) &= \text{exp}\left[\varepsilon\,\log(v)\,\mathbf{M_v^{(0)}}\right]\,\vec{g}_0(\varepsilon).
\end{align}
We computed numerically up to 120 digits the constants $\vec{g}_0(\varepsilon)$, through ${\cal{O}}(\varepsilon^6)$, by means of \texttt{DiffExp}. The solution of eqs.~(\ref{deq:v}) and (\ref{deq:x}) can be written in the form of generalised series expansion \cite{Caron-Huot:2017fxr,Abreu:2024xoh,Buccioni:2024gzo}
\begin{align}
\label{eq:solseries}
    \vec{f}(v,x) = T(\varepsilon,v,x)\,\text{exp}\left[\varepsilon\,\log(v)\,\mathbf{M_v^{(0)}}\right]\,\mathbb{P}\text{exp}\left[\varepsilon\int_1^x\,\mathbf{M_x}(v=0,t),dt\right]\,\vec{g}_0,
\end{align}
where the matrix $T(\varepsilon,v,x)$ has a Taylor expansion in $v$ and $\varepsilon$
\begin{equation}
\label{eq:Texpansion}
    T(\varepsilon,v,x) =  \mathbb{I} + \sum_{k,j\geq1} v^k\,\varepsilon^j\,T^{(k,j)}(x).
\end{equation}
The matrix $T(\varepsilon,v,x)$ controls the power corrections and it is not needed for the expansion of $\vec{f}(v,x)$ at leading power in $v$. However, in the quark-gluon amplitude (\ref{def:quaScat}), some of the 2dHPLs multiply rational factors involving spurious poles in $v$, thus requiring the subleading powers in eq.~(\ref{eq:Texpansion})
\begin{align}
    T^{(k,1)}(x) &= \frac{\mathbf{M_v^{(k)}}(x)}{k},\\
    T^{(k,j)}(x) &= \frac{1}{k}\left\{\big[\mathbf{M_v^{(0)}},T^{(k,\,j-1)}(x)\big] + \sum_{q=1}^{k-1} \mathbf{M_v^{(k-q)}}(x)T^{(q,\,j-1)}(x)\right\},\, \text{for } j>1.
\end{align}
The matrix $\mathbf{M_x}(v=0,x)$ in eq.~(\ref{eq:solseries}) has a very simple structure. Its entries are $\{1/x,1/(1+x)\}$, and the iterated integrals in eq.~(\ref{eq:solseries}) can be written in terms of HPLs, with letters $\{-1,0\}$, in the variable $x$. Finally, the numeric constants in $\vec{g}_0(\varepsilon)$, together with the constants arising in the path-ordered integrals of eq.~(\ref{eq:solseries}), can be written in terms of Riemann zeta values using the \texttt{FindIntegerNullVector} command in MATHEMATICA \cite{Mathematica}, which implements the PSLQ algorithm \cite{PSLQ}.    

By expanding the rational factors appearing in the amplitudes of ref.~\cite{Gehrmann:2023etk} and replacing the HPLs and 2dHPLs with their asymptotic expansion, eq.~(\ref{eq:solseries}), we obtain the Regge limit of the amplitudes for $q+g\to q+H$ and $g+g\to g+H$ up to ${\cal{O}}(\epsilon^4)$ at one loop and up to ${\cal{O}}(\epsilon^2)$ at two loops.
\subsection{The infrared factorisation in the Regge limit}
\label{sect:ir}
\input{IR}

\section{The Higgs impact factor at two loops}
\label{sec:IF2}

We are now in the position to match the amplitudes for Higgs + 3 partons with the asymptotic behaviour predicted by eq.~(\ref{Hggnnll}). We determine the Higgs impact factor at two loops, $c^{H(2)}$, by means of eq.~(\ref{eq:Hgggnnll2})
\begin{align}
\label{eq:extractIF2}
    c^{H(2)}&={\mathcal{M}}^{(2)}_{i\,g\to i\,H} - \frac{\left(\alpha^{(1)}\right)^2}{2}\,L^2-\left[\alpha^{(2)}+\alpha^{(1)}(c^{i(1)}+c^{H(1)})\right]\,L-c^{i(2)}\nonumber\\ 
      &-c^{\text{i}(1)}\,c^{H(1)}+\frac{\pi^2}{8}\left(\alpha^{(1)}\right)^2,
\end{align}
where the renormalised two-loop amplitudes, ${\mathcal{M}}^{(2)}_{i\,g\to i\,H}$, for $i=q,g$, were obtained in sec.~\ref{sec:M2Regge}, and the coefficients $\alpha^{(1)}$, $\alpha^{(2)}$ are given in eqs.~(\ref{alph}) and (\ref{alph2}), respectively. The impact factors $c^{i(p)}$, for $p=1,2$, and $c^{H(1)}$ are reported in appendix~\ref{app:IFcoef}. 
Thus, we extract the two-loop Higgs impact factor $c^{H(2)}(t,m_H^2,\tau)$ up to ${\cal{O}}(\epsilon^2)$. As a check on our calculation, we find that the impact factor does not change if we extract it either from the gluon amplitude, ${\cal{M}}_{g g\to g H}^{(2)}$ or from the quark amplitude ${\cal{M}}^{(2)}_{q g\to q H}$, thus verifying the universality of Regge factorisation, eq.~(\ref{Hggnnll}). 
It is convenient to use infrared factorisation to remove all the singularities of the impact factors. It is known \cite{DelDuca:2014cya,Falcioni:2021dgr} that the quark and gluon impact factors, up to two-loop order, are written as
\begin{equation}
    c^{i}(t,\tau) = Z_{\text{col}\,i}\left(\frac{t}{\mu^2},\alpha_s\right)\,\left(\frac{\tau}{-t}\right)^{\frac{\alpha_g(t)}{2}}\,\bar{D}_i(\alpha_s,t,\mu^2),
\end{equation}
where $Z_{\text{col}\,i}$ are defined in eq.~(\ref{eq:ZcolPart}) and the IR subtracted impact factors
\begin{equation}
\label{eq:Dbarexpansion}
    \bar{D}_i(\alpha_s,t,\mu^2) = 1+\sum_{n=1}^\infty\bar{D}_i^{(n)}(t,\mu^2)\,\left(\frac{\alpha_s}{4\pi}\right)^n
\end{equation}
are finite as $\epsilon\to 0$. Similarly, up to two-loop order, the Higgs impact factor is
\begin{align}
\label{eq:cHfact}
    c^{H}(t,m^2_{H},\tau)&=\frac{Z_{\text{col}\,gH}\left(\frac{m^2_{H\perp}}{\mu^2},\alpha_s\right)}{\cos\left(\frac{\pi}{2}\alpha_g(t)\right)}\,\left(\frac{\tau}{m_{H\perp}^2}\right)^\frac{\alpha_g(t)}{2}\bar{D}_H\left(\alpha_s,x,\mu^2\right),
\end{align}
where $x=-t/m_H^2$ and $m_{H\perp}^2=m_H^2\,(1+x)$. The factor $\cos(\pi/2\,\alpha_g(t))$, in the denominator of the equation above, can be absorbed by reorganising high-energy factorisation, eq.~(\ref{Hggnnll}), in terms of the symmetrised logarithm $L$ instead of using the sum $(s/\tau)^{\alpha_g}+(-s/\tau)^{\alpha_g}$. The contribution $\bar{D}_H\left(\alpha_s,x,\mu^2\right)$ is finite as $\epsilon\to 0$ through two loops. By expanding $\bar{D}_H(\alpha_s,x,\mu^2)$ as in eq.~(\ref{eq:Dbarexpansion}),
\begin{equation}
\label{eq:DbarHexpansion}
    \bar{D}_H(\alpha_s,x,\mu^2) = 1+\sum_{n=1}^\infty\bar{D}_H^{(n)}(x,\mu^2)\,\left(\frac{\alpha_s}{4\pi}\right)^n
\end{equation}
and setting $\mu^2=-t$, we get the one-loop coefficient
\begin{align}
\label{eq:DH1}
    \bar{D}^{(1)}_H\left(x\right)&=N_c\left[2\text{Li}_2(-x)-\log^2(x)+2\log(x)\log(1+x)+\frac{11}{2}\zeta_2+\frac{67}{18}-2i\pi\log\left(\frac{x}{1+x}\right)\right]\nonumber\\
    &-\frac{5}{9}n_f+{\cal{O}}(\eps),
\end{align}
in agreement with \cite{Celiberto:2022fgx,Fucilla:2024cpf}. At two loops, the impact factor structure is best shown by separating contributions of different transcendental weight. While in the ancillary files they are provided through ${\cal{O}}(\epsilon^2)$, for the sake of brevity here
we present the result up to ${\cal{O}}(\epsilon^0)$, which are of weight $4$ at most
\begin{align}
\label{eq:DH2}
    \bar{D}^{(2)}_H(x)&=\bar{D}^{(2)}_{H,w=4}(x)+\bar{D}^{(2)}_{H,\beta_0}(x)-6\zeta_3\left(N_c\,n_f+\frac{n_f}{N_c}\right)+N_c\,(67N_c-10n_f)\Bigg[\frac{\text{Li}_2(-x)}{3}\nonumber\\
    &-\frac{1}{6}\log^2(x)+\frac{1}{3}\log(x)\log(1+x)+\frac{17}{18}\zeta_2-\frac{i\pi}{3}\,\log\left(\frac{x}{1+x}\right)\Bigg]\nonumber\\
    &+N_c^2\Bigg[\frac{202}{27}\log(1+x)-\frac{202x^2+224x-122}{27(1+x)^2}\log(x)+\frac{48049x+51505}{648(1+x)}\nonumber\\
    &+i\pi\frac{4(5x+9)}{3(1+x)^2}\Bigg]-N_cn_f\Bigg[\frac{28}{27}\log(1+x)-\frac{28x^2+38x-107}{27(1+x)^2}\log(x)+\frac{16747x}{648(1+x)}\nonumber\\
    &+\frac{19555}{648(1+x)}+i\pi\frac{2x+15}{3(1+x)^2}\Bigg]+\frac{n_f}{N_c}\Bigg[\frac{\log(x)}{(1+x)^2}+\frac{55x+63}{8(1+x)}+\frac{i\pi}{(1+x)^2}\Bigg]+\frac{25}{54}n_f^2\nonumber\\
    &+{\cal{O}}(\epsilon),
\end{align}
where $\bar{D}^{(2)}_{w=4}(x)$ and $\bar{D}^{(2)}_{\beta_0}(x)$ are of weight $4$ and weight $3$, respectively. The former is
\begin{align}
    \bar{D}^{(2)}_{H,w=4}(x)&=
    8N_c^2\,\Bigg\{\text{Li}_4\left(\frac{x}{1+x}\right)-\frac{1}{2}\text{Li}_4(-x)+\frac{1}{2}\text{Li}_3(-x)\log(x)-\frac{1}{4}\text{Li}_2(-x)\log^2(x)\nonumber\\
    &+\frac{\log^4(x)}{16}+\frac{\log^4(1+x)}{24}+\frac{1}{4}\log^2(x)\log^2(1+x)-\frac{1}{4}\log^3(x)\log(1+x)\nonumber\\
    &-\frac{1}{6}\log(x)\log^3(1+x)+\zeta_2\Bigg(\frac{15}{8}\text{Li}_2(-x)-\frac{31}{16}\log^2(x)+\frac{31}{8}\log(x)\log(1+x)\nonumber\\
    &-\log^2(1+x)\Bigg)+\frac{\zeta_3}{8}\log\left(\frac{x}{1+x}\right)+\frac{277}{128}\zeta_4 + i\pi\,\Bigg[\frac{1}{2}\text{Li}_3(-x)\nonumber-\frac{1}{2}\text{Li}_2(-x)\log(x)\nonumber\\
    &+\frac{1}{4}\log^3(x)-\frac{3}{4}\log^2(x)\log(1+x)+\frac{1}{2}\log(x)\log^2(1+x)-\frac{1}{6}\log^3(1+x)\nonumber\\
    &-\frac{7}{8}\zeta_2\log\left(\frac{x}{1+x}\right)\Bigg]\Bigg\}.
\end{align}
We note that $\bar{D}^{(2)}_{{w}=4}(x)$ is written in terms of classical polylogarithms, in agreement with the observation that the finite remainder of the amplitude for Higgs + 3 gluons is expressed by means of this class of functions \cite{Duhr:2012fh}.
The contribution $\bar{D}^{(2)}_{\beta_0}(x)$ is proportional to the one-loop QCD beta function and it reads
\begin{align}
    \bar{D}^{(2)}_{H,\beta_0}(x)&=2N_c\,(11N_c-2n_f)\Bigg[-\text{Li}_3(-x)-\frac{1}{3}\text{Li}_3\left(\frac{1}{1+x}\right)+\frac{2}{3}\text{Li}_2(-x)\log(x)-\frac{\log^3(x)}{18}\nonumber\\
    &+\frac{1}{6}\log^2(x)\log(1+x)+\frac{\log^3(1+x)}{18}+\frac{\zeta_2}{3}\big(4\log(x)-5\log(1+x)\big)-\frac{13}{36}\zeta_3\nonumber\\
    &+i\pi\Bigg(\frac{2}{3}\text{Li}_2(-x)-\frac{\log^2(x)}{6}+\frac{1}{3}\log(x)\log(1+x)+\frac{\log^2(1+x)}{6}+\frac{2}{3}\zeta_2\Bigg)\Bigg].
\end{align}
Finally, we replace the two-loop Higgs impact factor obtained from eq.~(\ref{eq:extractIF2}) into eq.~(\ref{eq:Hgggnnll3}) and we get a prediction of the three-loop amplitude at NNLL. As a check, we verify that the infrared singularities match eq.~(\ref{eq:poles3loop}) through the linear terms in the high-energy logarithm $L$.

\section{Conclusions}
\label{sec:conc}

In the HEFT, we have considered the Regge limit of the two-loop amplitudes for Higgs boson production in association with a jet, expanded to NNLL accuracy.
We have shown that the contribution of the Regge cut can be set to zero at that accuracy, thanks to the simple colour structure of the amplitudes. Accordingly, in sec.~\ref{sec:IF2} we have determined for the first time the Higgs impact factor at two-loop accuracy in the HEFT, and based on that in eq.~(\ref{eq:Hgggnnll3}) we have predicted the Regge limit of the three-loop amplitudes for Higgs boson production in association with a jet, through the single-logarithmic term\footnote{After this work was released on \texttt{arxiv}, the three-loop amplitudes for Higgs+3 partons in the HEFT have been computed in ref.~\cite{Chen:2025utl}, but their Regge limit has not been carried out yet.}.

In planar ${\cal N} = 4$ SYM, the three-point form factor of the chiral stress-tensor multiplet is known through eight loops~\cite{Dixon:2020bbt,Dixon:2022rse}. By the principle of maximal transcendentality, it should be related to the highest weight part of the
HEFT amplitude for Higgs boson production in
association with a jet. We postpone exploring this relation to future work.

\section*{Acknowledgements}
We thank Michael Fucilla and Einan Gardi for useful discussions. GF is supported by the EU’s Horizon Europe research and innovation programme under the Marie Sk\l{}odowska-Curie grant 101104792, \texttt{QCDchallenge}.

\appendix


\section{Kinematics for Higgs  +  jet}
\label{sec:appa}

We consider the production of a parton of momentum $p_3$ and a Higgs boson of momentum $\ph$,
in the scattering between two partons of momenta $p_1$ and $p_2$, where 
all momenta are taken as outgoing.

Using light-cone coordinates $p^{\pm}= p_0\pm p_z $, and
complex transverse coordinates $p_{\perp} = p_x + i p_y$, with scalar
product,
\beq
2\, p\cdot q = p^+q^- + p^-q^+ - p_{\perp} q^*_{\perp} - p^*_{\perp} q_{\perp}\,,
\label{eq:scalprod}
\eeq
the four-momenta are
\begin{eqnarray}
p_1 &=& \left(p_1^-/2, 0, 0,-p_1^-/2 \right) 
     \equiv  \left(0, p_1^-; 0, 0 \right)\, ,\nonumber \\
p_2 &=& \left(p_2^+/2, 0, 0,p_2^+/2 \right) 
     \equiv  \left(p_2^+, 0; 0, 0\right)\, ,\nn\\
p_3 &=& \left( (p_3^+ + p_3^- )/2, 
                {\rm Re}[\pat],
                {\rm Im}[\pat], 
                (p_3^+ - p_3^- )/2 \right)\nonumber\\
    &\equiv& \left(|\pat| e^{y_3}, |\pat| e^{-y_3}; 
|\pat|\cos{\phi_3}, |\pat|\sin{\phi_3}\right) \, ,\nonumber \\
\ph &=& \left( (\ph^+ + \ph^- )/2, 
                {\rm Re}[\pht],
                {\rm Im}[\pht], 
                (\ph^+ - \ph^- )/2 \right)\nonumber\\
    &\equiv& \left(\mht e^{\yh}, \mht  e^{-\yh}; 
|\pht|\cos{\phih}, |\pht|\sin{\phih}\right) \,, \label{in}
\end{eqnarray}
where $y$ is the rapidity and $\mht = \sqrt{|\pht|^2+\mh^2}$ 
the Higgs transverse mass. 
The first notation in \eqn{in} is the standard representation 
$p^\mu =(p_0,p_x,p_y,p_z)$, while the second features light-cone
components, on which we have used the mass-shell conditions,
\begin{eqnarray}
0 &=& p_3^+ p_3^- - \pat \pat^\ast \,, \nn\\
\mh^2 &=& \php \phm - \pht \pht^\ast \,.
\label{eq:massshell}
\end{eqnarray}
Momentum conservation is
\begin{eqnarray}
0 &=& \pat +  \pht\, ,\nonumber \\
- p_2^+ &=& p_3^+ + \ph^+\, ,\label{nkin}\\ 
- p_1^- &=& p_3^- + \ph^-\, .\nonumber
\end{eqnarray}
Using momentum conservation, the mass-shell conditions (\ref{eq:massshell}) fulfil the constraint,
\beq
\mh^2 = \php \phm - p_3^+ p_3^- \,,
\eeq
and the Mandelstam invariants can be written as
\begin{eqnarray}
s_{12} &=& 2p_1\cdot p_2 =  (p_3^+ + \php) (p_3^- + \phm) \,,\nn\\
s_{23}  &=& 2p_2\cdot p_3 = -(p_3^+ + \php) p_3^-\,, \nn\\
s_{13}  &=& 2p_1\cdot p_3 = -p_3^+ (p_3^- + \phm) \,, \nn\\
s_{1 H} &=& (p_1+\ph)^2 = s_{23}\,, \nn\\
s_{2 H} &=& (p_2+\ph)^2 = s_{13}\,, \nn\\
s_{3 H} &=& (p_3+\ph)^2 = s_{12}\,.
\label{eq:mandinv}
\end{eqnarray}
Using momentum conservation (\ref{nkin}), the first of the equations above yields \eqn{Hjmtmcons}.

We use the following notation~\cite{Mangano:1990by} for
spinor products
\begin{equation}
\langle p k\rangle \equiv \langle p^- | k^+ \rangle\, , \qquad 
\left[pk\right] \equiv \langle p^+ | k^- \rangle\, , \qquad {\rm with}\;\;
\langle p k\rangle^* = {\rm sign}(p^0 k^0) \left[ k p\right]\, ,\label{spino}
\end{equation}
and currents
\beqa
\langle i| k | j\rangle &\equiv&
\langle i^-| \slash  \!\!\! k  |j^-\rangle = 
\langle i k \rangle \left[k j\right]\, ,\nonumber\\
\langle i| (k+l) | j\rangle &\equiv& 
\langle i^-| (\slash  \!\!\! k + \slash  \!\!\! l ) |j^-\rangle
= \langle i| k | j\rangle + \langle i| l | j\rangle\, 
,\label{currentsi}
\eeqa
and Mandelstam invariants
\begin{equation}
s_{pk} = 2\, p\cdot k = 
\langle p k \rangle \left[kp\right]\, . 
\end{equation}

Using the spinor representation of Ref.~\cite{DelDuca:1999ha},
the spinor products~(\ref{spino}) are
\begin{eqnarray}
\langle p_2 p_3\rangle &=& - i \sqrt{-p_2^+
\over p_3^+}\, \pat\, ,\label{spro}\\ 
\langle p_3 p_1\rangle &=&
i \sqrt{-p_1^- p_3^+}\, ,\nonumber\\ 
\langle p_2 p_1\rangle 
&=& -\sqrt{p_2^+p_1^-}\, ,\nonumber
\end{eqnarray}
where on $p_3$ we have used the mass-shell condition (\ref{eq:massshell}). 
The currents are obtained from \eqn{currentsi}.

\subsection{Regge limit}
\label{sec:mrk}

In the Regge limit, the light-cone momenta are strongly ordered,
\beq
\ph^+ \gg p_3^+ \, ,\qquad
|\pht| \simeq |\pat| \, .\label{mrk}
\eeq
Momentum conservation~(\ref{nkin}) becomes
\begin{eqnarray}
0 &=& p_{3_\perp} + \pht\, ,\nonumber \\
- p_2^+ &\simeq& p_H^+\, ,\label{mrkin}\\ 
- p_1^- &\simeq& p_3^-\, .\nonumber
\end{eqnarray}
To leading accuracy, the Mandelstam invariants (\ref{eq:mandinv}) are reduced to
\begin{eqnarray}
s_{12} &\simeq&  p_H^+ p_3^- \,,\nn\\
s_{23}  &\simeq& - p_H^+ p_3^- \,, \nn\\
s_{13}  &\simeq& -p_3^+ p_3^- \,.
\label{eq:kininvmrk}
\end{eqnarray}
\eqn{mrk} implies the hierarchy on the Mandelstam invariants,
\beq
s_{12} \gg - s_{13}\,.
\eeq
Introducing a parameter $\sigma$, the hierarchy above is equivalent to the rescaling, $s_{13} = \ord(\sigma)$.

\section{The amplitudes for Higgs $+$ three partons in the HEFT }
\label{sec:hjamp}

In the HEFT, where the loop-mediated Higgs-gluon coupling is replaced by an effective tree-level coupling,
the amplitude for Higgs $+$ three gluons, $p_1p_2\rightarrow p_3H$, 
can be written as
\beq
{\cal M}_{H3g}(p_1^{\nu_1}, p_2^{\nu_2}, p_3^{\nu_3}, \ph ) = \lambda\, \frac{g}2\,
(F^{a_3})_{a_1a_2}\, m_{H3g}(p_1^{\nu_1}, p_2^{\nu_2}, p_3^{\nu_3}, \ph )\,,
\label{eq:hggg0amp}
\eeq
with $\lambda$ as in \eqn{eq:heftcoupl}, and where colour matrices in the fundamental representation are normalised
as $\Tr (T^a T^b) = \delta^{ab}$, such that $[T^a, T^b] = (F^{b})_{ac} T^c$, with $(F^{b})_{ac} = i \sqrt{2} f^{abc}$.
The tree-level colour-ordered amplitudes are~\cite{Kauffman:1996ix}
\begin{eqnarray}
m^{(0)}_{H3g}(p_1^\oplus, p_2^\oplus, p_3^\oplus, \ph) &=& {\mh^4\over \la 1 2\ra \la 2 3\ra
   \la 3 1\ra} \, ,\label{ppp} \\  
m^{(0)}_{H3g}(p_1^\ominus,p_2^\oplus, p_3^\oplus, \ph) &=& {[23]^3 \over [1 2] [13]}\, ,
\label{mpp}
\end{eqnarray}
with spinor products and currents defined in app.~\ref{sec:appa}.
All of the other colour-ordered amplitudes can be obtained by relabelling and
by use of reflection symmetry, and parity inversion.
Parity inversion flips the helicities of all particles,
and it is accomplished by the substitution, 
$\la i j\ra \leftrightarrow [j i ]$.

\section{Anomalous dimensions}
\label{AppAnDim}

The perturbative expansion of the
cusp anomalous dimension~\cite{Kodaira:1982az,Korchemsky:1985xj,Moch:2004pa}, divided by the relevant quadratic Casimir factor $C_i$, is
\beq
\label{hatgammaK}
  \gamma_K (\as)  = \sum_{L=1}^\infty \gamma_K^{(L)} \left( \frac{\alpha_s}{\pi} \right)^L\,,
\eeq
with 
\beq
\gamma_K^{(1)} = 2\,, \qquad \gamma_K^{(2)} = \left( \frac{64}{18} + \frac{\delta_R}6 - \zeta_2 \right) C_A - \frac59 n_f\,.
\label{eq:k2}
\eeq
where
\begin{equation}
\delta_R = \left\{ \begin{array}{ll} 1 & \mbox{HV or CDR},\\
0 & \mbox{dimensional reduction}. \end{array} \right. \label{cp}
\end{equation}
The three-loop cusp anomalous dimension in HV and CDR scheme is
\begin{align}
\label{eq:k3}
    \gamma_K^{(3)}&=C_A^2\left(\frac{245}{48}-\frac{67}{18}\zeta_2+\frac{11}{12}\zeta_3+\frac{11}{4}\zeta_4\right)+C_An_f\left(-\frac{209}{216}+\frac{5}{9}\zeta_2-\frac{7}{6}\zeta_3\right)\nonumber\\
    &+C_Fn_f\left(-\frac{55}{48}+\zeta_3\right)-\frac{n_f^2}{54},
\end{align}
where $C_F=\frac{N_c^2-1}{2N_c}$. The perturbative expansion of the collinear anomalous dimension is
\beq
\label{collad}
  \gamma_i (\as)  = \sum_{L=1}^\infty \gamma_i^{(L)} \left( \frac{\alpha_s}{\pi} \right)^L\,, \qquad i = q, g\,,
\eeq
with the one-loop coefficients
\beq
\gamma_g^{(1)} = - \frac{\beta_0}{4}\,, \qquad \gamma_q^{(1)} = - \frac{3}4 C_F\,,
\label{eq:c1}
\eeq
where $\beta_0$ is the coefficient of the beta function \cite{Khriplovich:1969aa,Gross:1973id,Politzer:1973fx},
\beq
\beta_0 = \frac{11N_c - 2n_f}3\,.
\label{eq:b0k2}
\eeq
The two-loop anomalous dimensions are
\begin{align}
\label{eq:c2g}
    \gamma_g^{(2)}&=C_A^2\left(-\frac{173}{108}+\frac{11}{48}\zeta_2+\frac{\zeta_3}{8}\right)+C_An_f\left(\frac{8}{27}-\frac{\zeta_2}{24}\right)+\frac{C_Fn_f}{8},\\
\label{eq:c2q}    
    \gamma_q^{(2)}&=C_F^2\left(-\frac{3}{32}+\frac{3}{4}\zeta_2-\frac{3}{2}\zeta_3\right)+C_AC_F\left(-\frac{961}{864}-\frac{11}{16}\zeta_2+\frac{13}{8}\zeta_3\right)\nonumber\\
    &+C_Fn_f\left(\frac{65}{432}+\frac{\zeta_2}{8}\right).
\end{align}
Note that, as customary in the literature, the expansion in \Eqns{hatgammaK}{collad} is in $\alpha_s/\pi$,
while the impact factor (\ref{fullv}) and the Regge trajectory (\ref{alphb}) are expanded in $\alpha_s/4\pi$.

The integrals of the anomalous dimensions defined in eqs.~(\ref{Kdef})-(\ref{Bdef}) are expanded as follows
\begin{equation}
    K\left(\alpha_s\right)=\sum_{n\geq1}K^{(n)}\,\left(\frac{\alpha_s}{4\pi}\right)^n,
\end{equation}
and similarly for $B_i(\alpha_s)$ and $D(\alpha_s)$. The coefficients $K^{(i)}$ read
\begin{align}
\label{eq:K1}
   K^{(1)}&=\frac{\gamma_K^{(1)}}{\epsilon},\\
\label{eq:K2}   
   K^{(2)}&=-\frac{\gamma_K^{(1)}\beta_0}{2\epsilon^2}+\frac{2\gamma_K^{(2)}}{\epsilon},\\
\label{eq:K3}   
   K^{(3)}&=\frac{\gamma_K^{(1)}\beta_0^2}{3\epsilon^3}-\frac{\gamma_K^{(1)}\beta_1-4\beta_0\gamma_K^{(2)}}{3\epsilon^2}+\frac{16\gamma_K^{(3)}}{3\epsilon},
\end{align}
where the two-loop beta function is \cite{Caswell:1974gg,Jones:1974mm,Egorian:1978zx}
\begin{equation}
\label{eq:b1}
    \beta_1=\frac{34C_A^2}{3}-\frac{10}{3}C_An_f-2C_Fn_f.
\end{equation}
The coefficients $B_i^{(n)}$ are obtained from eqs.~(\ref{eq:K1})-(\ref{eq:K3}), with the replacement $\gamma_K^{(n)}\to2\gamma_i^{(n)}$, which follows from the factor of 2 in the definition of $B_i(\alpha_s)$, eq.~(\ref{Idef}), compared to eq.~(\ref{Kdef}). The coefficients of $D(\alpha_s)$ are
\begin{align}
\label{eq:D1}
    D^{(1)}&=-\frac{\gamma_K^{(1)}}{\epsilon^2},\\
\label{eq:D2}    
    D^{(2)}&=\frac{3\gamma_K^{(1)}\beta_0}{4\epsilon^3}-\frac{\gamma_K^{(2)}}{\epsilon^2},\\
\label{eq:D3}    
    D^{(3)}&=-\frac{11}{18}\frac{\gamma_K^{(1)}\beta_0^2}{\epsilon^4}+\frac{4\gamma_K^{(1)}\beta_1+10\gamma_K^{(2)}\beta_0}{9\epsilon^3}-\frac{16\gamma_K^{(3)}}{9\epsilon^2}.
\end{align}
Using the expressions in eqs.~(\ref{eq:K1})-(\ref{eq:K3}) and (\ref{eq:D1})-(\ref{eq:D3}), we get the coefficients $Z^{(i)}$ of the infrared operator defined in eq.~(\ref{eq:IRfact})
\begin{align}
    Z^{(1)}&=Z_{\text{col},i}^{(1)}+Z_{\text{col},gH}^{(1)}+\widetilde{Z}^{(1)},\\
    Z^{(2)}&=Z_{\text{col},i}^{(2)}+Z_{\text{col},gH}^{(2)}+\widetilde{Z}^{(2)}+Z_{\text{col},i}^{(1)}Z_{\text{col},gH}^{(1)}+Z_{\text{col},i}^{(1)}\widetilde{Z}^{(1)}+Z_{\text{col},gH}^{(1)}\widetilde{Z}^{(1)},\\
    Z^{(3)}&=Z_{\text{col},i}^{(3)}+Z_{\text{col},gH}^{(3)}+\widetilde{Z}^{(3)}+Z_{\text{col},i}^{(2)}Z_{\text{col},gH}^{(1)}+Z_{\text{col},i}^{(1)}Z_{\text{col},gH}^{(2)}+Z_{\text{col},i}^{(1)}\widetilde{Z}^{(2)}\nonumber\\
    &+Z_{\text{col},i}^{(2)}\widetilde{Z}^{(1)}+Z_{\text{col},i}^{(1)}\widetilde{Z}^{(2)}+Z_{\text{col},i}^{(1)}Z_{\text{col},gH}^{(1)}\widetilde{Z}^{(1)},
\end{align}
where, following from the definitions in eqs.~(\ref{eq:Ztilde}), (\ref{eq:ZcolPart}) and (\ref{eq:ZcolH}), the coefficients ${\widetilde{Z}}^{(i)}$, for $i=1,2,3$ are
\begin{align}
\label{eq:Ztilde1}
{\widetilde{Z}}^{(1)}&=K^{(1)}\,\widetilde{L},\\
\label{eq:Ztilde2}
{\widetilde{Z}}^{(2)}&=\frac{1}{2}\left(K^{(1)}\,\widetilde{L}\right)^2+K^{(2)}\widetilde{L},\\
\label{eq:Ztilde3}
{\widetilde{Z}}^{(3)}&=\frac{1}{6}\left(K^{(1)}\,\widetilde{L}\right)^3+K^{(1)}K^{(2)}\widetilde{L}^2+K^{(3)}\widetilde{L},
\end{align}
and 
\begin{align}
\label{eq:ZcolH1}
    Z_{\text{col}\,gH}^{(1)}&= \frac{C_AK^{(1)}}{2}\log\left(\frac{m_{H\perp}^2}{\mu^2}\right)+\frac{C_AD^{(1)}}{2}+B_g^{(1)},\\
\label{eq:ZcolH2}    
    Z_{\text{col}\,gH}^{(2)}&= \frac{\left(C_AK^{(1)}\right)^2}{8}\log^2\left(\frac{m_{H\perp}^2}{\mu^2}\right)+\log\left(\frac{m_{H\perp}^2}{\mu^2}\right)\left[\frac{K^{(2)}}{2}+\frac{K^{(1)}(D^{(1)}+2B_g^{(1)})}{4}\right]\nonumber\\
    &+\frac{1}{8}\left(D^{(1)}+2B^{(1)}_g\right)^2+\frac{D^{(2)}+2B^{(2)}_g}{2}.
\end{align}
The coefficients $Z_{\text{col}\,i}^{(n)}$, with $n=1,2$ are obtained from eqs~(\ref{eq:ZcolH1}) and (\ref{eq:ZcolH2}) with the replacements $C_A\to\mathbf{{T}}^2_i$, $K^{(i)}\to2K^{(i)}$ and $B_g^{(n)}\to2B^{(n)}_i$. $Z^{(3)}_{\text{col}\,H}$ and $Z^{(3)}_{\text{col}\,i}$ enter only to N\textsuperscript{3}LL accuracy.
\section{Impact factors}
\label{app:IFcoef}
We begin this appendix by providing the terms of higher order in $\epsilon$ in the one-loop impact factors, which are required in eqs.~(\ref{eq:Hgggnnll2}) and (\ref{eq:Hgggnnll3}). It is convenient to extract the dependence on the factorisation scale $\tau$, which is controlled by eq.~(\ref{Hggnll}) to all orders in $\alpha_s$
\begin{equation}
    c^{i}(\xi_i,\tau')=c^{i}(\xi_i,\tau)\left(\frac{\tau'}{\tau}\right)^{\frac{\alpha_g(t)}{2}},
\end{equation}
where $\xi_i$ labels the remaining arguments of the impact factor $c_i$ for $i=q,g,H$. The dependence on the scale $\mu^2$ is given by the renormalisation of the couplings $\alpha_s$ and $\lambda$, with~\cite{Gehrmann:2023etk}
\begin{align}
    \lambda_{\text{bare}} &= \lambda\left[1-\frac{\beta_0}{\epsilon}\left(\frac{\alpha_s}{4\pi}\right)+\left(\frac{\beta_0^2}{\epsilon^2}-\frac{\beta_1}{\epsilon}\right)\left(\frac{\alpha_s}{4\pi}\right)^2+{\cal{O}}(\alpha_s^3)\right],
\end{align}
where $\beta_0$ and $\beta_1$ are given in eqs.~(\ref{eq:b0k2}) and (\ref{eq:b1}), respectively. For the impact factors of the quark and of the gluon we get  
\begin{align}
\label{eq:IF1loopGEN}
    c^{i(1)}(t,\tau,\mu^2)&=\left(\frac{\mu^2}{-t}\right)^\epsilon\left[\bar{c}^{i(1)}+\frac{\beta_0}{2\epsilon}+\frac{\bar{\alpha}^{(1)}}{2}\log\left(\frac{\tau}{-t}\right)\right]-\frac{\beta_0}{2\epsilon},
\end{align}
where we use the notation
\begin{align}
    \bar{\alpha}^{(p)} &= \alpha^{(p)}(t,\mu^2=-t),\\
    \bar{c}^{i(p)}&=c^{i(p)}(t,\tau=-t,\mu^2=-t).
\end{align}
The coefficient $\bar{\alpha}^{(1)}$ is read off eq.~(\ref{alph}). We obtain the one-loop gluon impact factor in the HV scheme by replacing in the equation above
\begin{align}
\label{eq:IF1loopG}
\bar{c}^{g(1)}&=\kappa_\Gamma\Bigg[-\frac{2C_A}{\epsilon^2}-\frac{\beta_0}{\epsilon}+C_A\left(-\frac{67}{18}+3\zeta_2\right)+\frac{5}{9}n_f+\epsilon\left(C_A\left(-\frac{202}{27}-\frac{11}{12}\zeta_2+\zeta_3\right)\right.\nonumber\\
&\left.+n_f\left(\frac{28}{27}+\frac{\zeta_2}{6}\right)\right)+\epsilon^2\left(C_A\left(-\frac{1214}{81}-\frac{77}{18}\zeta_3+3\zeta_4\right)+n_f\left(\frac{164}{81}+\frac{7}{9}\zeta_3\right)\right)\nonumber\\
&+\epsilon^3\left(C_A\left(-\frac{7288}{243}-\frac{209}{32}\zeta_4+\zeta_5\right)+n_f\left(\frac{976}{243}+\frac{19}{16}\zeta_4\right)\right)+{\cal{O}}(\epsilon^4)\Bigg]
\end{align}
while the one-loop quark impact factor in the HV scheme is given by
\begin{align}
\label{eq:IF1loopQ}
\bar{c}^{q(1)}&=\kappa_\Gamma\Bigg[-\frac{2C_F}{\epsilon^2}-\frac{3C_F}{\epsilon}+C_A\left(\frac{85}{18}+3\zeta_2\right)-8C_F-\frac{5}{9}n_f+\epsilon\left(C_A\left(\frac{256}{27}-\frac{11}{12}\zeta_2+\zeta_3\right)\right.\nonumber\\
&\left.-16C_F+n_f\left(-\frac{28}{27}+\frac{\zeta_2}{6}\right)\right)+\epsilon^2\left(C_A\left(\frac{1538}{81}-\frac{77}{18}\zeta_3+3\zeta_4\right)-32C_F\right.\nonumber\\
&\left.+n_f\left(-\frac{164}{81}+\frac{7}{9}\zeta_3\right)\right)+\epsilon^3\left(C_A\left(\frac{9232}{243}-\frac{209}{32}\zeta_4+\zeta_5\right)-64C_F\right.\nonumber\\
&\left.+n_f\left(-\frac{976}{243}+\frac{19}{16}\zeta_4\right)\right)+{\cal{O}}(\epsilon^4)\Bigg]
\end{align}
The one-loop Higgs impact factor is given by
\begin{align}
\label{eq:IFHiggs1GEN}
    c^{H(1)}(t,m_H^2,\tau,\mu^2)&=\left(\frac{\mu^2}{-t}\right)^\epsilon\Bigg[\bar{c}^{H(1)}(x)+\frac{\beta_0}{\epsilon}+\frac{\bar{\alpha}^{(1)}}{2}\log\left(\frac{\tau}{m_{H\perp}^2}\right)\Bigg]-\frac{\beta_0}{\epsilon},
\end{align}
where $x=\frac{-t}{m_H^2}$. We use the bar notation to indicate the Higgs impact factor at $p$-loops evaluated at fixed values of the renormalisation and of the factorisation scales. We choose $\bar{c}^{H(p)}(x)=c^{H(p)}(t,m_H^2,\tau=m_{H\perp}^2,\mu^2=-t)$, with
\begin{align}
\label{eq:IFHiggs1loop}
\bar{c}^{H(1)}(x)&=\kappa_\Gamma\Bigg[-\frac{C_A}{\epsilon^2}-\frac{\beta_0}{2\epsilon}+\frac{C_A}{\epsilon}\Big(G(-1,x)-G(0,x)\Big)+C_A\left(\frac{67}{18}+5\zeta_2+2G(-1,0,x)\right.\nonumber\\
&\left.-2G(0,0,x)+2i\pi\Big(G(-1,x)-G(0,x)\Big)\vphantom{\frac{1}{2}}\right)-\frac{5}{9}n_f+\epsilon\left(C_A\left(\frac{148+202x}{27\,(1+x)}\right.\right.\nonumber\\
&+\frac{2x\,G(0,x)}{(1+x)^2}-\frac{11}{6}\zeta_2+6\zeta_2\Big(G(0,x)-G(-1,x)\Big)+2G(-1,0,0,x)-2G(0,0,0,x)\nonumber\\
&\left.\left.+\zeta_3+2i\pi\,\Big(\frac{x}{(1+x)^2}+G(-1,0,x)-G(0,0,x)+\zeta_2\Big)\right)+n_f\left(-\frac{28}{27}+\frac{\zeta_2}{3}\right)\right)\nonumber\\
&+\epsilon^2\left(C_A\left(\frac{728+1214x}{81\,(1+x)}+\frac{4x-2}{(1+x)^2}G(0,x)+\frac{2x}{(1+x)^2}\Big(G(0,0,x)-3\zeta_2\Big)-\frac{77}{9}\zeta_3\right.\right.\nonumber\\
&+6\zeta_2\Big(G(0,0,x)-G(-1,0,x)\Big)+2G(-1,0,0,0,x)-2G(0,0,0,0,x)-\frac{17}{2}\zeta_4\nonumber\\
&+2i\pi\left(\frac{2x-1}{(1+x)^2}+\frac{x\,G(0,x)}{(1+x)^2}+\zeta_2\Big(G(0,x)-G(-1,x)\Big)+G(-1,0,0,x)\right.\nonumber\\
&\left.\left.\left.-G(0,0,0,x)\vphantom{\frac{1}{2}}\right)\right)+n_f\left(-\frac{164}{81}+\frac{14}{9}\zeta_3\right)\right)+\epsilon^3\left(C_A\left(\frac{3886+7288x}{243(1+x)}\right.\right.\nonumber\\
&+\frac{8x-6}{(1+x)^2}G(0,x)+\frac{2(1-2x)}{(1+x)^2}\Big(3\zeta_2-G(0,0,x)\Big)+\frac{2x}{(1+x)^2}\Big(G(0,0,0,x)\nonumber\\
&-3\zeta_2G(0,x)\Big)-\frac{209}{16}\zeta_4+6\zeta_2\Big(G(0,0,0,x)-G(-1,0,0,x)\Big)-\frac{15}{2}\zeta_4\Big(G(0,x)\nonumber\\
&-G(-1,x)\Big)+2G(-1,0,0,0,0,x)-2G(0,0,0,0,0,x)+\zeta_5+2i\pi\left(\frac{4x-3}{(1+x)^2}\right.\nonumber\\
&-\frac{1-2x}{(1+x)^2}G(0,x)+\frac{x}{(1+x)^2}\Big(G(0,0,x)-\zeta_2\Big)+\zeta_2\Big(G(0,0,x)-G(-1,0,x)\Big)\nonumber\\
&\left.\left.\left.+G(-1,0,0,0,x)-G(0,0,0,0,x)-\frac{3}{4}\zeta_4\right)\right)+n_f\left(-\frac{976}{243}+\frac{19}{8}\zeta_4\right)\right)+{\cal{O}}(\epsilon^4)\Bigg],
\end{align}
in terms of the Goncharov multiple polylogarithms $G$ \cite{goncharov2011multiplepolylogarithmscyclotomymodular}. The two-loop impact factors of the quark and of the gluon, which contribute to eqs.~(\ref{eq:Hgggnnll2}) and (\ref{eq:Hgggnnll3}), are derived from the results of ref.~\cite{Falcioni:2021dgr}. By using the same notation as in eq.~(\ref{eq:IF1loopGEN}), we write  
\begin{align}
\label{eq:IF2loopGEN}
    c^{i(2)}(t,\tau,\mu^2)&=\left(\frac{\mu^2}{-t}\right)^{2\epsilon}\Bigg[\bar{c}^{i(2)}+\frac{1}{2}\log\left(\frac{\tau}{-t}\right)\,(\bar{\alpha}^{(2)}+\bar{c}^{i(1)}\bar{\alpha}^{(1)})+\frac{(\bar{\alpha}^{(1)})^2}{8}\log^2\left(\frac{\tau}{-t}\right)\nonumber\\
    &+\frac{3\beta_0^2}{8\epsilon^2}+\frac{3\beta_0}{4\epsilon}\left(2\bar{c}^{i(1)}+\bar{\alpha}^{(1)}\log\left(\frac{\tau}{-t}\right)\right)+\frac{\beta_1}{4\epsilon}\Bigg]-\frac{3\beta_0}{4\epsilon}\left(\frac{\mu^2}{-t}\right)^\epsilon\Bigg[\frac{\beta_0}{\epsilon}+2\bar{c}^{i(1)}\nonumber\\
    &+\bar{\alpha}^{(1)}\log\left(\frac{\tau}{-t}\right)\Bigg]+\frac{3\beta_0^2}{8\epsilon^2}-\frac{\beta_1}{4\epsilon},
\end{align}
where the two-loop gluon impact factor is obtained by replacing $\bar{c}^{g(1)}$ from eq.~(\ref{eq:IF1loopG}) and
\begin{eqnarray}
\allowdisplaybreaks
\label{eq:IF2loopG}
\bar{c}^{g(2)} &=& \kappa_\Gamma^2\Bigg[\frac{2C_A^2}{\epsilon^4}+\frac{7C_A\beta_0}{2\epsilon^3}+\frac{1}{\epsilon^2}\left(C_A^2\Big(\frac{103}{6}-5\zeta_2\Big)-\frac{49}{9}C_An_f+\frac{4}{9}n_f^2\right)+\frac{1}{\epsilon}\left(C_A^2\Big(\frac{853}{54}\right.\nonumber\\
&-&\left.\frac{11}{6}\zeta_2-\zeta_3\Big)+C_An_f\left(-\frac{38}{9}+\frac{\zeta_2}{3}\right)+C_Fn_f+\frac{10}{27}n_f^2\right)+C_A^2\left(\frac{10525}{648}+\frac{1033}{36}\zeta_2\right.\nonumber\\
&+&\left.\frac{121}{3}\zeta_3-\frac{55}{4}\zeta_4\right)+C_An_f\left(-\frac{452}{81}-\frac{58}{9}\zeta_2-\frac{10}{3}\zeta_3\right)+C_Fn_f\left(\frac{55}{12}-4\zeta_3\right)\nonumber\\
&+&n_f^2\left(\frac{29}{54}+\frac{\zeta_2}{3}\right)+\epsilon\left(C_A^2\Big(-\frac{24191}{648}+\frac{9895}{216}\zeta_2+\frac{452}{3}\zeta_3+\frac{4895}{48}\zeta_4+\zeta_2\zeta_3-41\zeta_5\Big)\right.\nonumber\\
&+&C_An_f\Big(\frac{973}{972}-\frac{236}{27}\zeta_2-\frac{254}{9}\zeta_3-\frac{301}{24}\zeta_4\Big)+C_Fn_f\Big(\frac{1711}{72}+\frac{\zeta_2}{2}-\frac{38}{3}\zeta_3-6\zeta_4\Big)\nonumber\\
&+&\left.n_f^2\Big(\frac{188}{243}+\frac{5}{18}\zeta_2+\frac{14}{9}\zeta_3\Big)\right)+{\cal{O}}(\epsilon^2)\Bigg]
\end{eqnarray}
Similarly, we get the two-loop quark impact factor by replacing $\bar{c}^{q(1)}$, eq.~(\ref{eq:IF1loopQ}), into eq.~(\ref{eq:IF2loopGEN}) and
\begin{eqnarray}
\allowdisplaybreaks
\label{eq:IF2loopQ}
\bar{c}^{q(2)}&=&\kappa_\Gamma^2\Bigg[\frac{2C_F^2}{\epsilon^4}+\frac{C_F}{\epsilon^3}\left(\frac{3}{2}\beta_0+6C_F\right)+\frac{1}{\epsilon^2}\left(\frac{41}{2}C_A^2+C_AC_F\Big(-\frac{23}{3}-5\zeta_2\Big)+\frac{2}{3}C_Fn_f\right)\nonumber\\
&+&\frac{1}{\epsilon}\left(C_F^2\Big(\frac{221}{4}+6\zeta_2-12\zeta_3\Big)+C_AC_F\Big(-\frac{1513}{36}-\frac{43}{6}\zeta_2+11\zeta_3\Big)+C_Fn_f\Big(\frac{89}{18}-\frac{\zeta_2}{3}\Big)\!\right)\nonumber\\
&+&C_F^2\left(\frac{1151}{8}+29\zeta_2-30\zeta_3-22\zeta_4\right)+C_AC_F\left(-\frac{40423}{216}-\frac{1447}{36}\zeta_2+84\zeta_3+\frac{43}{2}\zeta_4\right)\nonumber\\
&+&C_Fn_f\left(\frac{530}{27}+\frac{29}{18}\zeta_2-2\zeta_3\right)+C_A^2\left(\frac{13195}{216}+\frac{73}{2}\zeta_2-\frac{43}{3}\zeta_3-\frac{53}{4}\zeta_4\right)\nonumber\\
&+&C_An_f\Big(-\frac{385}{27}-5\zeta_2-\frac{14}{3}\zeta_3\Big)+\frac{25}{54}n_f^2+\epsilon\left(C_F^2\Big(\frac{5741}{16}+\frac{217}{2}\zeta_2-184\zeta_3-39\zeta_4\right.\nonumber\\
&+&\vphantom{\frac{1}{2}}16\zeta_2\zeta_3-12\zeta_5\Big)+C_AC_F\Big(-\frac{844711}{1296}-\frac{7315}{54}\zeta_2+\frac{7639}{18}\zeta_3+\frac{2043}{16}\zeta_4-56\zeta_2\zeta_3\nonumber\\
&-&111\zeta_5\Big)+C_Fn_f\Big(\frac{5137}{81}+\frac{517}{54}\zeta_2-\frac{5}{9}\zeta_3-\frac{9}{8}\zeta_4\Big)+C_A^2\Big(\frac{184255}{648}+\frac{4525}{72}\zeta_2-\frac{2233}{18}\zeta_3\nonumber\\
&+&\frac{77}{8}\zeta_4+41\zeta_2\zeta_3+82\zeta_5\Big)+C_An_f\Big(-\frac{19999}{324}-\frac{89}{18}\zeta_2-\frac{34}{9}\zeta_3-\frac{55}{4}\zeta_4\Big)\nonumber\\
&+&\left.n_f^2\Big(\frac{140}{81}-\frac{5}{18}\zeta_2\Big)\right)+{\cal{O}}(\epsilon^2)\Bigg]
\end{eqnarray}
Finally, the two-loop impact factor of the Higgs is given by
\begin{align}
\label{eq:IFHiggs2GEN}
    c^{H(2)}(t,m_H^2,\tau,\mu^2)&=\left(\frac{\mu^2}{-t}\right)^{2\epsilon}\Bigg[\bar{c}^{H(2)}(x)+\frac{1}{2}\log\left(\frac{\tau}{m_{H\perp}^2}\right)\left(\bar{\alpha}^{(2)}+2\bar{\alpha}^{(1)}\frac{\beta_0}{\epsilon}+\bar{\alpha}^{(1)}\bar{c}^{H(1)}(x)\right)\nonumber\\
    &+\frac{\left(\bar{\alpha}^{(1)}\right)^2}{8}\log^2\left(\frac{\tau}{m_{H\perp}^2}\right)+\frac{2\beta_0}{\epsilon}\bar{c}^{H(1)}(x)+\frac{\beta_0^2}{\epsilon^2}+\frac{\beta_1}{\epsilon}\Bigg]-\left(\frac{\mu^2}{-t}\right)^\epsilon\Bigg[\frac{2\beta_0^2}{\epsilon^2}\nonumber\\
    &+\frac{\beta_0}{\epsilon}\left(2\bar{c}^{H(1)}(x)+\bar{\alpha}^{(1)}\log\left(\frac{\tau}{m_{H\perp}}\right)\right)\Bigg]+\frac{\beta_0^2}{\epsilon^2}-\frac{\beta_1}{\epsilon},
\end{align}
with $\bar{c}^{H(1)}(x)$ written in eq.~(\ref{eq:IFHiggs1loop}) and
\begin{eqnarray}
    \bar{c}^{H(2)}(x)&=&\kappa_\Gamma^2\Bigg[\frac{C_A^2}{2\epsilon^4}+\frac{C_A}{\epsilon^3}\left(\frac{5\beta_0}{4}+C_A\Big(G(0,x)-G(-1,x)\Big)\right)+\frac{1}{\epsilon^2}\left(C_A^2\Big(-\frac{13}{24}\right.\nonumber\\
    &+&\frac{11}{3}\big(G(0,x)-G(-1,x)\big)+G(-1,-1,x)-3G(-1,0,x)-G(0,-1,x)\nonumber\\
    &+&3G(0,0,x)-\frac{3}{2}\zeta_2+2i\pi\big(G(0,x)-G(-1,x)\big)\Big)+C_An_f\Big(-1+\frac{2}{3}G(-1,x)\nonumber\\
    &-&\left.\frac{2}{3}G(0,x)\Big)+\frac{n_f^2}{6}\right)+\frac{1}{\epsilon}\left(C_A^2\bigg(-\frac{2021+2237x}{108\,(1+x)}+\frac{67}{9}G(-1,x)\nonumber\right.\\
    &-&\frac{67+152x+67x^2}{9\,(1+x)^2}G(0,x)-\frac{11}{3}(G(-1,0,x)-G(0,0,x)+\zeta_2)\nonumber\\
    &-&10\zeta_2\big(G(0,x)-G(-1,x)\big)+4G(-1,-1,0,x)+2G(-1,0,-1,x)\nonumber\\
    &-&8G(-1,0,0,x)-4G(0,-1,0,x)-2G(0,0,-1,x)+8G(0,0,0,x)-\frac{\zeta_3}{2}\nonumber\\
    &+&2i\pi\Big(-\frac{x}{(1+x)^2}+\frac{11}{6}(G(0,x)-G(-1,x))+2G(-1,-1,x)-3G(-1,0,x)\nonumber\\
    &-&2G(0,-1,x)+3G(0,0,x)-\zeta_2\Big)\bigg)+C_An_f\bigg(\frac{121}{27}+\frac{10}{9}\big(G(0,x)-G(-1,x)\big)\nonumber\\
    &+&\frac{2}{3}\big(G(-1,0,x)-G(0,0,x)+\zeta_2\big)-2i\pi\frac{G(0,x)-G(-1,x)}{3}\bigg)+\frac{C_Fn_f}{2}\nonumber\\
    &-&\left.\frac{5}{27}n_f^2\right)+C_A^2\Bigg(\frac{39169+29449x}{648\,(1+x)}+\frac{350+754x+404x^2}{27\,(1+x)^2}G(-1,x)\nonumber\\
    &+&\frac{28-781x-404x^2}{27\,(1+x)^2}G(0,x)+\frac{2x\,G(0,-1,x)}{(1+x)^2}+\frac{67+140x+67x^2}{3\,(1+x)^2}G(-1,0,x)\nonumber\\
    &-&\frac{67+152x+67x^2}{3\,(1+x)^2}G(0,0,x)+\frac{1193+2494x+1193x^2}{18\,(1+x)^2}\zeta_2+\frac{22}{3}G(-1,-1,0,x)\nonumber\\
    &+&\frac{11}{3}G(-1,0,0,x)\!-\!\frac{22}{3}G(0,-1,0,x)\!-\!\frac{11}{3}G(0,0,0,x)+22(G(0,x)-G(-1,x))\zeta_2\nonumber\\
    &+&\frac{77}{18}\zeta_3-8G(-1,-1,-1,0,x)+12G(-1,-1,0,0,x)+10G(-1,0,-1,0,x)\nonumber\\
    &+&2G(-1,0,0,-1,x)\!-\!22G(-1,0,0,0,x)\!+\!8G(0,-1,-1,0,x)\!-\!2G(0,0,0,-1,x)\nonumber\\
    &-&12G(0,-1,0,0,x)-10G(0,0,-1,0,x)+22G(0,0,0,0,x)+4\zeta_2\big(7G(0,-1,x)\nonumber\\
    &+&12G(-1,0,x)-7G(-1,-1,x)-12G(0,0,x)\big)+\frac{41}{2}\zeta_4+2i\pi\Big(\frac{14-x}{2\,(1+x)^2}\nonumber\\
    &+&\left(\frac{67}{6}+\frac{x}{(1+x)^2}\right)G(-1,x)-\left(\frac{67}{6}+\frac{2x}{(1+x)^2}\right)G(0,x)+\frac{11}{3}G(-1,-1,x)\nonumber\\
    &+&\frac{11}{6}G(-1,0,x)-\frac{11}{3}G(0,-1,x)-\frac{11}{6}G(0,0,x)+\frac{11}{2}\zeta_2-4G(-1,-1,-1,x)\nonumber\\
    &+&6G(-1,-1,0,x)+5G(-1,0,-1,x)-10G(-1,0,0,x)+4G(0,-1,-1,x)\nonumber\\
    &-&6G(0,-1,0,x)-5G(0,0,-1,x)+10G(0,0,0,x)-5\zeta_2(G(0,x)-G(-1,x))\Big)\!\Bigg)\nonumber\\
    &+&C_An_f\Bigg(-\frac{2681+2033x}{162\,(1+x)}-\frac{56}{27}G(-1,x)+\frac{4(14x^2+28x-13)}{27\,(1+x)^2}G(0,x)\nonumber\\
    &-&\frac{10}{3}G(-1,0,x)+\frac{10}{3}G(0,0,x)-\frac{34}{3}\zeta_2-\frac{4}{3}G(-1,-1,0,x)-\frac{2}{3}G(-1,0,0,x)\nonumber\\
    &+&\frac{4}{3}G(0,-1,0,x)+\frac{2}{3}G(0,0,0,x)-4(G(0,x)-G(-1,x))\zeta_2-\frac{115}{9}\zeta_3\nonumber\\
    &+&2i\pi\Big(\!\!-\!\frac{2}{(1+x)^2}+\frac{5}{3}(G(0,x)-G(-1,x))-\frac{2}{3}G(-1,-1,x)-\frac{1}{3}G(-1,0,x)\nonumber\\
    &+&\frac{2}{3}G(0,-1,x)+\frac{1}{3}G(0,0,x)-\zeta_2\Big)\Bigg)+C_Fn_f\left(-\frac{63+55x}{4(1+x)}-\frac{2G(0,x)}{(1+x)^2}+12\zeta_3\right.\nonumber\\
    &-&\left.\frac{2i\pi}{(1+x)^2}\right)+n_f^2\left(\frac{19}{162}+\frac{2}{9}\zeta_2\right)+{\cal{O}}(\epsilon)\Bigg]
\end{eqnarray}

\bibliographystyle{JHEP}
\bibliography{refs.bib}

\end{document}

%% file: IR.tex
The analysis of the amplitudes and the matching with the asymptotic formula in the Regge limit, eq.~(\ref{Hggnnll}), is streamlined by using input from the infrared factorised expression \cite{Akhoury:1978vq,Sen:1982bt,Kidonakis:1998nf,Catani:1998bh,Sterman:2002qn,Becher:2009cu,Gardi:2009qi,Ma:2019hjq}
\beq
  {\cal M} \left(\frac{p_i}{\mu}, \alpha_s (\mu^2)
  \right) \, = \, Z \left(\frac{p_i}{\mu}, \alpha_s(\mu^2) \right) \, \, 
  {\cal H} \left(\frac{p_i}{\mu}, \alpha_s(\mu^2) \right) \, ,
\label{Mfac}
\eeq
where the $Z$ operator captures the infrared poles and ${\cal H}$ the finite part of the amplitude.
The former is universal and it is given by 
\begin{equation}
\label{def:Zop}
    Z\left(\frac{p_i}{\mu}, \alpha_s(\mu^2) \right) = \mathbb{P}\text{exp}\left[-\frac{1}{2}\int_0^{\mu^2}\frac{d\lambda^2}{\lambda^2}\mathbf{\Gamma}_n\left(\{s_{ij}\},\alpha_s(\lambda^2),\lambda^2\right)\right],
\end{equation}
where $\mathbf{\Gamma}_n$ is the anomalous dimension for $n$ external (massless) quarks and gluons. This quantity has been calculated at three loops \cite{Almelid:2015jia,Almelid:2017qju}. At four-loop order, for $n=3$ external partons, the anomalous dimension can be written in the form \cite{Gardi:2009qi,Becher:2019avh}
\begin{align}
\label{eq:Gamma}
\mathbf{\Gamma}_3\left(\{s_{ij}\},\alpha_s(\lambda^2),\lambda^2\right) &= \mathbf{\Gamma}^{\text{dip}}_3\left(\{s_{ij}\},\alpha_s(\lambda^2),\lambda^2\right) + \mathbf{\Gamma}_{3,\text{4T-3L}}\left(\{s_{ij}\},\alpha_s(\lambda^2),\lambda^2\right) \nn\\
& + \mathbf{\Gamma}_{3, \text{Q4T-2,3L}}\left(\{s_{ij}\},\alpha_s(\lambda^2),\lambda^2\right) + {\cal{O}}(\alpha_s^5).
\end{align}
In the equation above $\mathbf{\Gamma}^{\text{dip.}}_3$ is the sum over dipoles contribution \cite{Becher:2009cu,Becher:2009qa,Gardi:2009qi,Gardi:2009zv}
\begin{align}
\label{eq:gammaDip}
    \mathbf{\Gamma}^{\text{dip}}_3\left(\{s_{ij}\},\alpha_s(\lambda^2),\lambda^2\right)&=-\frac{1}{2}\widehat{\gamma}_K(\alpha_s)\Big[\log\left(\frac{s_{12}e^{-i\,\pi}}{\lambda^2}\right)\,\mathbf{T_1}\cdot\mathbf{T_2}+\log\left(\frac{-s_{13}}{\lambda^2}\right)\,\mathbf{T_1}\cdot\mathbf{T_3}\nn\\
    &+\log\left(\frac{-s_{23}}{\lambda^2}\right)\,\mathbf{T_2}\cdot\mathbf{T_3}\Big] + \sum_{i=1}^3 \gamma_{J_i}(\alpha_s(\lambda^2)),
\end{align}
where $\widehat{\gamma}_K$ is the universal cusp anomalous dimension and $\gamma_{J_i}$  is the 
anomalous dimension for the partonic jet function $J_i$, see app.~\ref{AppAnDim}. The second term in eq.~(\ref{eq:Gamma}) is 
\begin{align}
    \mathbf{\Gamma}_{3,\text{4T-3L}}\left(\{s_{ij}\},\alpha_s(\lambda^2),\lambda^2\right)&=f(\alpha_s)\sum_{i\neq j\neq k=1}^3\mathbf{\cal{T}}_{iijk},
\end{align}
where $\mathbf{\cal{T}}_{iijk}=f^{ade}f^{bce}\,\frac{1}{2}\left(\mathbf{T}_i^a\mathbf{T}_i^b+\mathbf{T}_i^a\mathbf{T}_i^b\right)\,\mathbf{T}_j^c\,\mathbf{T}_k^d$. The coefficient $f(\alpha_s)$ has been computed to three loops~\cite{Almelid:2015jia}
\begin{equation}
    f(\alpha_s)=\left(\frac{\alpha_s}{\pi}\right)^3\,\frac{\zeta_5+2\zeta_2\zeta_3}{4}+{\cal{O}}(\alpha_s^4).
\end{equation}
The last term in eq.~(\ref{eq:Gamma}) is
\begin{align}
    \mathbf{\Gamma}_{3, \text{Q4T-2,3L}}\left(\{s_{ij}\},\alpha_s(\lambda^2),\lambda^2\right)&=-\frac{1}{2}\sum_R g_R(\alpha_s)\left[\sum_{i\neq j=1}^3({\cal{D}}^R_{iijj}+2{\cal{D}}^R_{iiij})\log\frac{s_{ij}e^{-i\pi}}{\lambda^2}\right.\nonumber\\
    &\left.+\sum_{i\neq j\neq k=1}^3{\cal{D}}^R_{ijkk}\log\frac{s_{ij}e^{-i\pi}}{\lambda^2}\right],
\end{align}
where ${\cal{D}}^R_{ijkl}=\frac{1}{4!}\sum_{\sigma\in{\cal{S}}_n}\text{Tr}\left[T_R^{\sigma(a)}T_R^{\sigma(b)}T_R^{\sigma(c)}T_R^{\sigma(d)}\right]\,\mathbf{T}_i^a\mathbf{T}_j^b\mathbf{T}_k^c\mathbf{T}_l^d$. The coefficient $g_R(\alpha_s)={\cal{O}}(\alpha_s^4)$ is the contribution of the quartic Casimir invariants to the cusp anomalous dimension and it was computed in \cite{Huber:2019fxe,Henn:2019swt,vonManteuffel:2020vjv}. 

We take the Regge limit of eq.~(\ref{eq:Gamma}), thus we can replace $s_{23}$ with $-s_{12}$
up to power--suppressed corrections, app.~\ref{sec:mrk}. At NNLL accuracy, only the term in eq.~(\ref{eq:gammaDip}) contributes. Indeed, $\mathbf{\Gamma}_{3,\text{4T-3L}}$, which is ${\cal{O}}(\alpha_s^3)$, does not depend on any kinematic variable, thus it contributes to N\textsuperscript{3}LL accuracy. Similarly, $\mathbf{\Gamma}_{3, \text{Q4T-2,3L}}$ is of ${\cal{O}}(\alpha_s^4)$ and contains at most a single logarithm of $s_{12}$. The limit of the sum over dipoles in eq.~(\ref{eq:gammaDip}) is evaluated by following the steps described in ref.~\cite{DelDuca:2011ae}. Introducing the scale integrals,
\beqa
  K \Big(\alpha_s (\mu^2) \Big) & \equiv &
  - \frac14 \int_0^{\mu^2} \frac{d \lambda^2}{\lambda^2} \, 
  \widehat{\gamma}_K \left(\alpha_s(\lambda^2) 
  \right) \, , \label{Kdef} \\
  D \Big(\alpha_s (\mu^2) \Big) & \equiv &
  - \frac14 \int_0^{\mu^2} \frac{d \lambda^2}{\lambda^2} \, 
  \widehat{\gamma}_K \left(\alpha_s(\lambda^2) \right)
  \ln \left( \frac{\mu^2}{{\lambda^2}} \right) \, ,
  \label{Idef} \\
  B_i \Big(\alpha_s (\mu^2) \Big) & \equiv &
  - \frac12 \int_0^{\mu^2} \frac{d\lambda^2}{\lambda^2} \, 
  \gamma_{J_i} \left(\alpha_s (\lambda^2) \right) \,,
  \label{Bdef}
\eeqa
of the cusp and collinear anomalous dimensions, which are reported in appendix~\ref{AppAnDim}, we write the $Z$ operator in factorised form,
\begin{equation}
\label{eq:IRfact}
Z \left( \frac{p_i}{\mu}, \alpha_s \right)\, = \widetilde{Z}\left( \frac{s_{12}}{\sqrt{- s_{13}}\, \mht}, \alpha_s \right) \,  Z_{\rm{col}\,i} \left( \frac{s_{13}}{\mu^2}, \alpha_s(\mu^2) \right)\,Z_{\rm{col}\, gH} \left( \frac{\mht^2}{\mu^2}, \alpha_s(\mu^2) \right),
\end{equation}
where $\widetilde{Z}$ encodes the dependence on high-energy logarithms which are accompanied by infrared poles
\begin{equation}
\label{eq:Ztilde}
 \widetilde{Z}\left( \frac{s_{12}}{\sqrt{- s_{13}}\, \mht}, \alpha_s \right) = \exp\left[K\left(\alpha_s(\mu^2)\right)\,C_A\,\widetilde{L}\,\right],
\end{equation}
with the signature-even logarithm $\widetilde{L}$ defined as follows
\begin{equation}
 \widetilde{L}=\left[ \ln \left(\frac{s_{12}}{\sqrt{- s_{13}}\, \mht} \right) -i\frac{\pi}{2}\right] = \frac{1}{2}\left[\ln\left(\frac{s_{12}+i0}{\sqrt{- s_{13}}\, \mht} \right) + \ln\left(\frac{-s_{12}-i0}{\sqrt{- s_{13}}\, \mht} \right)\right].
\end{equation}
In eq.~(\ref{eq:IRfact}), the factors $Z_{\rm{col}\,i}$ and $Z_{\rm{col}\,gH}$, which do not depend on the high-energy logarithm, generate the collinear divergences associated to the massless outgoing particles
\beqa
\label{eq:ZcolPart}
Z_{\rm{col}\,i} \left( \frac{s_{13}}{\mu^2}, \alpha_s \right) &=&
\exp \left[2 B_i \left(\alpha_s \right) 
 +  D \left(\alpha_s \right) {\bf T}_i^2
   + K \left(\alpha_s \right) \, 
   \ln \left(\frac{- s_{13}}{\mu^2} \right) {\bf T}_i^2 \right] \,, \\
\label{eq:ZcolH}
   Z_{\rm{col}\,gH} \left( \frac{\mht^2}{\mu^2}, \alpha_s \right) &=&
\exp \left[ B_{\rm g} \left(\alpha_s \right) 
 + \frac12 D \left(\alpha_s \right) C_A
   + \frac12 K \left(\alpha_s \right) \, 
   \ln \left(\frac{\mht^2}{\mu^2} \right) C_A \right] \,.
\eeqa
The structure of the infrared poles generated by eq.~(\ref{eq:IRfact}) is compatible with the high-energy factorisation in eq.~(\ref{Hggnnll}). The collinear factors in eqs.~(\ref{eq:ZcolPart}) and (\ref{eq:ZcolH}) are naturally associated with the singularities of the impact factors~\cite{DelDuca:2013ara,DelDuca:2014cya,Caron-Huot:2013fea} of the parton $i$ and of the Higgs, respectively, as discussed in sec.~\ref{sec:IF2}. The high-energy logarithms exponentiate according to eq.~(\ref{eq:Ztilde}). Indeed, the integral $K(\alpha_s)$ of the universal cusp anomalous dimension provides the IR poles of the renormalised gluon Regge trajectory, as shown to hold at two loops~\cite{Korchemskaya:1996je}, at three loops~\cite{Falcioni:2021dgr} and conjecturally to all orders~\cite{Falcioni:2021buo}. Notably, at NNLL accuracy the operator $\tilde{Z}$ doesn't include any term that is associated to the Regge cut in eq.~(\ref{Hggnnll}). Therefore, the IR structure justifies the prescription in eq.~(\ref{eq:cut2;;3}) for the Regge cut through three loops and it suggests that $\hat{M}^{(n)}_{\text{cut}}=0$ to NNLL accuracy.

Eq.~(\ref{eq:IRfact}) is consistent with the expression of the $Z$ operator for $2\to2$ coloured partons \cite{DelDuca:2011ae}, if the latter is expanded in terms of the signature-symmetric logarithm in eq.~(\ref{eq:SymLog2to2}). However, in the case of four coloured partons, $\widetilde{Z}$ acquires an imaginary part, proportional to $\mathbf{T}^2_s-\mathbf{T}^2_u$, which breaks the Regge pole behaviour of the amplitudes \cite{DelDuca:2011ae} and is consistent with IR poles of the Regge cut \cite{DelDuca:2013ara,DelDuca:2014cya,Falcioni:2021buo}. With only three partons, colour conservation imposes $\mathbf{T}^2_s=\mathbf{T}^2_u$, thus removing the source of imaginary parts and the need for a Regge cut contribution at the level of IR poles.

Eq.~(\ref{eq:IRfact}) correctly predicts the poles of the amplitudes in sec.~\ref{sec:M2Regge}, which can be written in terms of finite remainders
\begin{align}
    {\mathcal{H}}^{(1)}_{ig\to iH}&={\mathcal{M}}^{(1)}_{ig\to iH}-Z^{(1)}{\mathcal{M}}^{(0)}_{ig \to iH},\\
    {\mathcal{H}}^{(2)}_{ig\to iH}&={\mathcal{M}}^{(2)}_{ig\to iH}-Z^{(2)}{\mathcal{M}}^{(0)}_{ig \to iH}-Z^{(1)}{\mathcal{H}}^{(1)}_{ig\to iH},
\end{align}
where $Z=\sum Z^{(n)}(\alpha_s/(4\pi))^n$ is readily computed using expansions of $K(\alpha_s)$, $B_i(\alpha_s)$ and $D(\alpha_s)$ in appendix~\ref{AppAnDim}. Using the finite remainders ${\cal{H}}^{(1)}_{ig \to iH}$ to ${\cal{O}}(\epsilon^4)$ and ${\cal{H}}^{(2)}_{ig \to iH}$ to ${\cal{O}}(\epsilon^2)$ we get the poles of the three-loop amplitudes
\begin{equation}
\label{eq:poles3loop}
    {\mathcal{M}}^{(3)}_{ig\to iH}=Z^{(3)}{\mathcal{M}}^{(0)}_{ig \to iH}+Z^{(2)}{\mathcal{H}}^{(1)}_{ig \to iH} + Z^{(1)}{\mathcal{H}}^{(2)}_{ig\to iH}+{\cal{O}}(\epsilon^0).
\end{equation}